\documentclass[%
 aip,
 amsmath,amssymb,
 preprint,
]{revtex4-1}

\usepackage{graphicx}
\usepackage{dcolumn}
\usepackage{bm}
\usepackage[utf8]{inputenc} 
\usepackage[T1]{fontenc}
\usepackage{CJKutf8}

\usepackage{mathrsfs}

\usepackage{etoolbox}

\usepackage[colorlinks=true, citecolor=red]{hyperref}
\begin{document}
	\preprint{AIP/123-QED}
	\title{\Large Long-term predictions of turbulence by implicit U-Net enhanced Fourier neural operator}
	\author{Zhijie Li(\begin{CJK}{UTF8}{gbsn}李志杰\end{CJK})} %
	\affiliation{Department of Mechanics and Aerospace Engineering, Southern University of Science and Technology, Shenzhen 518055, China}
	\affiliation{Guangdong-Hong Kong-Macao Joint Laboratory for Data-Driven Fluid Mechanics and Engineering Applications, Southern University of Science and Technology, Shenzhen 518055, China}
	
	\author{Wenhui Peng(\begin{CJK}{UTF8}{gbsn}彭文辉\end{CJK})} %
	\affiliation{Department of Mechanics and Aerospace Engineering, Southern University of Science and Technology, Shenzhen 518055, China}
	\affiliation{Guangdong-Hong Kong-Macao Joint Laboratory for Data-Driven Fluid Mechanics and Engineering Applications, Southern University of Science and Technology, Shenzhen 518055, China}
	
	\author{Zelong Yuan(\begin{CJK}{UTF8}{gbsn}袁泽龙\end{CJK})} %
	\affiliation{Department of Mechanics and Aerospace Engineering, Southern University of Science and Technology, Shenzhen 518055, China}
	\affiliation{Guangdong-Hong Kong-Macao Joint Laboratory for Data-Driven Fluid Mechanics and Engineering Applications, Southern University of Science and Technology, Shenzhen 518055, China}
	
	\author{Jianchun Wang(\begin{CJK}{UTF8}{gbsn}王建春\end{CJK})}%
	\homepage{wangjc@sustech.edu.cn}
	\affiliation{Department of Mechanics and Aerospace Engineering, Southern University of Science and Technology, Shenzhen 518055, China}
	\affiliation{Guangdong-Hong Kong-Macao Joint Laboratory for Data-Driven Fluid Mechanics and Engineering Applications, Southern University of Science and Technology, Shenzhen 518055, China}


\begin{abstract}
Long-term predictions of nonlinear dynamics of three-dimensional (3D) turbulence are very challenging for machine learning approaches. In this paper, we propose an implicit U-Net enhanced Fourier neural operator (IU-FNO) for stable and efficient predictions on the long-term large-scale dynamics of turbulence. The IU-FNO model employs implicit recurrent Fourier layers for deeper network extension and incorporates the U-net network for the accurate prediction on small-scale flow structures. The model is systematically tested in large-eddy simulations of three types of 3D turbulence, including forced homogeneous isotropic turbulence (HIT), temporally evolving turbulent mixing layer, and decaying homogeneous isotropic turbulence. The numerical simulations demonstrate that the IU-FNO model is more accurate than other FNO-based models including vanilla FNO, implicit FNO (IFNO) and U-Net enhanced FNO (U-FNO), and dynamic Smagorinsky model (DSM) in predicting a variety of statistics including the velocity spectrum, probability density functions (PDFs) of vorticity and velocity increments, and instantaneous spatial structures of flow field. Moreover, IU-FNO improves long-term stable predictions, which has not been achieved by the previous versions of FNO. Besides, the proposed model is much faster than traditional LES with DSM model, and can be well generalized to the situations of higher Taylor-Reynolds numbers and unseen flow regime of decaying turbulence.

\end{abstract}

\maketitle
\section{Introduction}
Neural networks (NNs) have been widely applied to improve or replace the conventional modeling of turbulent flows in computational fluid dynamics (CFD).\cite{brunton2020machine,duraisamy2019turbulence,liu2020deep,vignon2023recent,zuo2023fast} Various strategies based on NNs have been developed to enhance Reynolds-averaged Navier-Stokes simulation (RANS) and large-eddy simulation (LES) of turbulence.\cite{maulik2018data,yuan2021dynamic,wang2021artificial,beck2018deep,xie2019artificial,wang2018investigations,sarghini2003neural,gamahara2017searching,zhou2019subgrid,li2021data,guan2022stable,ling2016reynolds,tabe2023priori} Beck et al. proposed convolutional neural networks (CNNs) and residual neural networks (RNNs) to construct accurate subgrid-scale (SGS) models for LES.\cite{beck2019deep} Zhou et al. used an artificial neural network to develop a new SGS model for LES of isotropic turbulent flows.\cite{zhou2019subgrid} Park and Wang also applied NNs to learn closures of SGS stress and thus improve the accuracy of turbulence modeling.\cite{park2021toward,wang2021artificial} Yang et al. introduced several physical insights to improve the extrapolation capabilities of neural networks for LES wall modeling.\cite{yang2019predictive} 

Deep neural networks have demonstrated a remarkable performance in approximating highly non-linear functions.\cite{lecun2015deep} Several recent studies have focused on approximating the complete Navier-Stokes equations using deep neural networks.\cite{lusch2018deep,sirignano2018dgm,tang2021exploratory,kovachki2023neural,goswami2022deep} Once trained, ``black-box'' neural network models can rapidly make inferences on modern computers, and can be much more efficient than traditional CFD methods. Moreover, some researchers have explored incorporating additional physical knowledge into deep learning methods.\cite{cai2021physics,wang2020towards,lanthaler2022error,karniadakis2021physics,raissi2017physics} Raissi et al. introduced a physics-informed neural networks (PINN) to solve general nonlinear partial differential equations.\cite{raissi2019physics} Xu et al. utilized the physics-informed deep learning to address the missing flow dynamics by treating the governing equations as a parameterized constraint.\cite{xu2021explore} Chen et al. proposed a theory-guided hard constraint projection method to convert governing equations into a form that is easy to handle through discretization and then implements hard constraint optimization through projection in a patch.\cite{chen2021theory} Wang et al. incorporated physical constraints into the neural network design and developed a turbulent flow network (TF-Net). The TF-Net offers the flexibility of the learned representations and achieves the state-of-the-art prediction accuracy.\cite{wang2020towards} Jin et al. developed the Navier-Stokes flow nets (NSFnets) by embedding the governing equations, initial conditions, and boundary conditions into the loss function.\cite{jin2021nsfnets}  

While most previous neural network architectures are good at learning mappings between finite-dimensional Euclidean spaces, they are limited in their generalization ability for different parameters, initial conditions or boundary conditions.\cite{raissi2019physics,wu2020data,xu2021deep,lu2022comprehensive,oommen2022learning} Recently, Li et al. proposed a novel Fourier neural operator (FNO) framework capable of efficiently learning the mapping between infinite dimensional spaces from input-output pairs.\cite{li2020fourier} The FNO model outperforms current state-of-the-art models, including U-Net,\cite{chen2019u} TF-Net,\cite{wang2020towards} and ResNet,\cite{he2016deep} in two-dimensional (2D) turbulence prediction. Peng et al. presented an FNO model coupled with the attention that can effectively reconstruct statistical properties and instantaneous flow structures of 2D turbulence at high Reynolds numbers.\cite{peng2022attention} Wen et al. proposed an U-net enhanced FNO (U-FNO) for solving multiphase flow problems with superior accuracy and efficiency.\cite{wen2022u} You et al. developed an implicit Fourier neural operator (IFNO), to model the increment between layers as an integral operator to capture the long-range dependencies in the feature space.\cite{you2022learning} The developments and applications of FNO-based models have been increasing,\cite{li2022fouriergeo,jiang2023fourier,tran2021factorized,renn2023forecasting,li2021physics,guibas2021adaptive,hao2023gnot,benitez2023fine,choubineh2023fourier} however, the majority of the works have been focused on one-dimensional (1D) and two-dimensional (2D) problems. Modeling 3D turbulence using deep neural networks is a greater challenge due to the significant increase in the size and dimension of simulation data compared to 2D problems.\cite{momenifar2022dimension} Moreover, modeling the non-linear interactions in 3D turbulence demands significant model complexity and a large number of parameters. Training models with such a huge number of parameters can be computationally expensive and requires significant memory usage, which can be very challenging due to hardware limitations. 

Recently, Mohan et al. developed two reduced models of 3D homogeneous isotropic turbulence (HIT) and scalar turbulence based on the deep learning methods including convolutional generative adversarial network (C-GAN) and compressed convolutional long-short-term-memory (CC-LSTM) network.\cite{mohan2020spatio} Ren et al. proposed a data-driven model for predicting turbulent flame evolution based on machine learning methods with long short-term memory (LSTM) and convolutional neural network-long short-term memory (CNN-LSTM). The CNN-LSTM model has been shown to outperform the LSTM model in terms of overall performance.\cite{ren2021predictive} Nakamura et al. combined a 3D convolutional neural network autoencoder (CNN-AE) and a long short-term memory (LSTM) to predict the 3D channel flow.\cite{nakamura2021convolutional} Lehmann et al. applied the FNO to predict ground motion time series from a 3D geological description.\cite{lehmann2023fourier} Li et al. utilized FNO for large-eddy simulation (LES) of 3D turbulence.\cite{li2022fourier} Peng et al. proposed a linear attention coupled Fourier neural operator (LAFNO) for the simulation of 3D isotropic turbulence and free-shear turbulence.\cite{peng2023linear} In this work, we propose an implicit U-Net enhanced Fourier neural operator (IU-FNO) as a surrogate model for LES of turbulence, in order to achieve stable, efficient and accurate predictions on the long-term large-scale dynamics of turbulence. 

The rest of the paper is organized as follows: Section \ref{sec:2} describes the governing equations of the large-eddy simulation and three classical subgrid-scale models. Section \ref{sec:3} introduces three types of previous Fourier neural operator architectures, including vanilla FNO, implicit FNO (IFNO) and U-Net enhanced FNO (U-FNO). In Section \ref{sec:4}, we propose a new FNO-based model, namely IU-FNO model. Section \ref{sec:5} introduces the data generation and training process, and presents the a $posteriori$ performance of the IU-FNO model in comparison to other FNO-based models and classical dynamic Smagorinsky model for three types of turbulent flows, including the forced homogeneous isotropic turbulence (HIT), free-shear mixing layer turbulence, and decaying homogeneous isotropic turbulence. Discussions and conclusions are finally drawn in Section \ref{sec:6} and \ref{sec:7} respectively.

\section{\label{sec:2}Governing equations and subgrid scale model}
This section provides a brief introduction to the filtered incompressible Navier-Stokes (NS) equations for classical LES models for the unclosed subgrid-scale (SGS) stress.

The governing equations of the three-dimensional incompressible turbulence are given by \cite{pope2000,Ishihara2009}
	\begin{equation}
	\frac{\partial u_i}{\partial x_i}=0,
	\label{eq1}
	\end{equation}
	\begin{equation}
	\frac{\partial u_i}{\partial t}+\frac{\partial\left(u_i u_j\right)}{\partial x_j}=-\frac{\partial p}{\partial x_i}+v \frac{\partial^2 u_i}{\partial x_j \partial x_j}+\mathcal{F}_i.
	\label{eq2}
	\end{equation}
Here $u_i$ denotes the $i$-th component of velocity, $p$ is the pressure divided by the constant density, $v$ represents the kinematic viscosity, and $\mathcal{F}_i$ stands for a large-scale forcing to the momentum of the fluid in the $i$-th coordinate direction. In this paper, the convention of summation notation is employed. 

The kinetic energy $E_k$ is defined as $E_k=\int_0^{\infty}E(k)dk=\frac{1}{2}\left(u^{rms}\right)^2$, where $E(k)$ is the energy spectrum, and $u^{r m s}=\sqrt{\left\langle u_i u_i\right\rangle}$ is the root mean square (rms) of the velocity, and $\langle\cdot\rangle$ denotes a spatial average along the homogeneous direction. In addition, the Kolmogorov length scale $\eta$, the Taylor length scale $\lambda$, and the Taylor-scale Reynolds number $Re_\lambda$ are defined, respectively, as \cite{wyp2022,pope2000}
	\begin{equation}
	\eta=\left(\frac{\nu^3}{\varepsilon}\right)^{1 / 4}, \quad \lambda=\sqrt{\frac{5 \nu}{\varepsilon}} u^{r m s}, \quad Re_\lambda=\frac{u^{r m s} \lambda}{\sqrt{3} \nu},
	\label{eq3}
	\end{equation}
where $\varepsilon=2 v\left\langle S_{i j} S_{i j}\right\rangle$ denotes the average dissipation rate and $S_{i j}=\frac{1}{2}\left(\partial u_i / \partial x_j+\partial u_j / \partial x_i\right)$ represents the strain rate tensor. Furthermore, the integral length scale $L_I$ and the large-eddy turnover time $\tau$ are given by\cite{pope2000}
	\begin{equation}
	L_I = \frac{3 \pi}{2\left(u^{r m s}\right)^2} \int_0^{\infty} \frac{E(k)}{k} d k,\quad \tau=\frac{L_I}{u^{r m s}}.
	\label{eq4}
	\end{equation}

A filtering methodology can be implemented to decompose the physical variables of turbulence into distinct large-scale and sub-filter small-scale components.\cite{lesieur1996,meneveau2000} The filtering operation is defined as $\bar{f}(\mathbf{x})=\int_\Omega f(\mathbf{x}-\mathbf{r}) G(\mathbf{r}, \mathbf{x} ; \Delta) d \mathbf{r}$, where $f$ represents a variable in physical space, and $\Omega$ is the entire domain. $G$ and $\Delta$ are the filter kernel and filter width, respectively.\cite{pope2000,sagaut2006} For any variable $f$ in Fourier space, a filtered variable is given by $\bar{f}(\boldsymbol{k})=\hat{G}(\boldsymbol{k}) f(\boldsymbol{k})$. In the present study, a sharp spectral filter $\hat{G}(\boldsymbol{k})=$ $H\left(k_c-|\boldsymbol{k}|\right)$ is utilized in Fourier space for homogeneous isotropic turbulence.\cite{pope2000} Here, the cutoff wavenumber $k_c=\pi / \Delta$, and $\Delta$ denotes the filter width. The Heaviside step function $H(x)=1$ if $x \geq 0$; otherwise $H(x)=0$.\cite{pope2000,chang2022} 

The filtered incompressible Navier-Stokes equations can be derived for the resolved fields as follows \cite{pope2000,sagaut2006}
	\begin{equation}
	\frac{\partial \bar{u}_i}{\partial x_i}=0,	
	\label{eq5}
	\end{equation}
	\begin{equation}
	\frac{\partial \bar{u}_i}{\partial t}+\frac{\partial\left(\bar{u}_i \bar{u}_j\right)}{\partial x_j}=-\frac{\partial \bar{p}}{\partial x_i}-\frac{\partial \tau_{i j}}{\partial x_j}+v \frac{\partial^2 \bar{u}_i}{\partial x_j \partial x_j}+\overline{\mathcal{F}}_i .
	\label{eq6}
	\end{equation}
Here, $\tau_{ij}$ is the unclosed sub-grid scale (SGS) stress defined by $\tau_{i j}=\overline{u_i u_j}-\bar{u}_i \bar{u}_j$. In order to solve the LES equations, it is crucial to model the SGS stress as a function of the filtered variables.

Subgrid-scale (SGS) models have been developed for the unclosed terms in the filtered incompressible Navier-Stokes equations. These models aim to accurately capture the nonlinear interactions between the resolved large-scales and unresolved small-scales.\cite{moser2021,johnson2022} Appendix~\ref{app:LES} provides a comprehensive introduction to three classical LES models, including dynamic Smagorinsky model (DSM), velocity gradient model (VGM) and dynamic mixed model (DMM).

\section{\label{sec:3}Related neural operator and modified methods}
Compared with traditional numerical methods and other neural operator methods, FNO shows a strong adaptability and generalization in dealing with high-dimensional and large-scale data.\cite{li2022fourier,guibas2021adaptive,rashid2022learning,pathak2022fourcastnet,li2022fouriergeo} The main idea of FNO is to use Fourier transform to map high-dimensional data into the frequency domain, and approximate nonlinear operators by learning the relationships between Fourier coefficients through neural networks. FNO can learn the rule of an entire family of PDE.\cite{li2022fourier} This part will mainly introduce the FNO and some typical improved methods based on it, including U-Net enhanced FNO (U-FNO) and implicit Fourier neural operator(IFNO).

\subsection{\label{sec:3.1}The Fourier neural operator}
\begin{figure*}
	\includegraphics[width=1\linewidth]{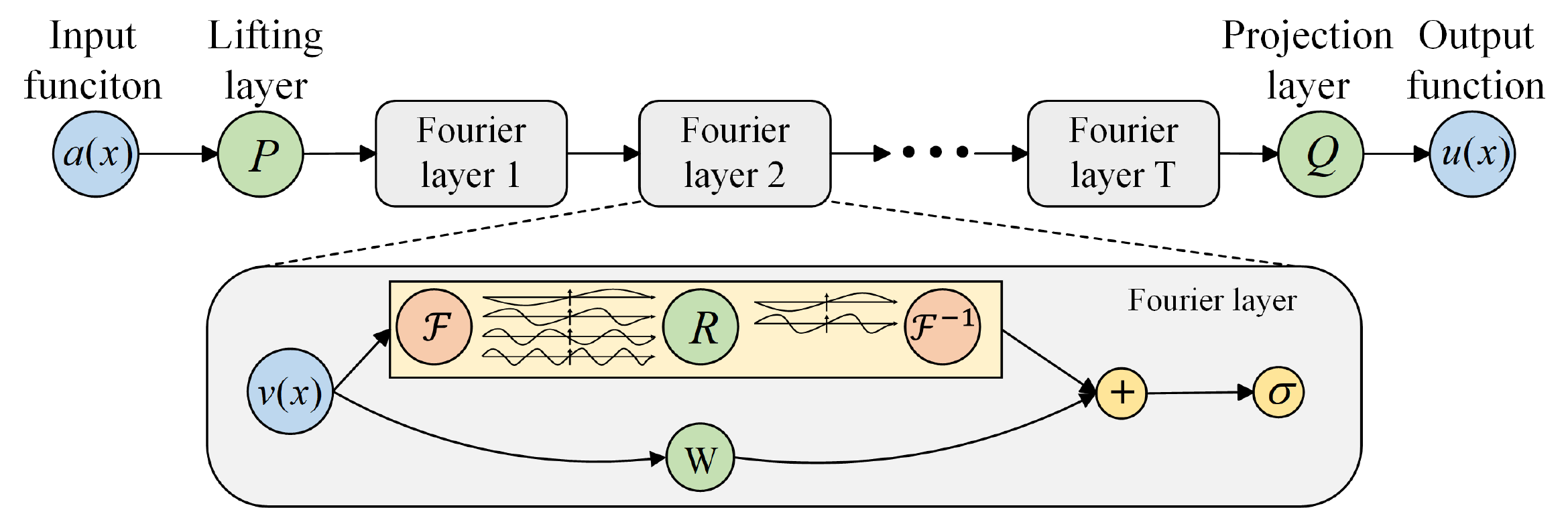}
	\caption{The Fourier neural operator (FNO) architecture.}
	\label{NNFNO}
	\end{figure*}

The Fourier neural operators (FNO) aims to map between two infinite-dimensional spaces by training on a finite set of input-output pairs. Denote $D \subset \mathbb{R}^d$ as a bounded, open set and $\mathcal{A}=\mathcal{A}\left(D ; \mathbb{R}^{d_a}\right)$ and $\mathcal{U}=\mathcal{U}\left(D ; \mathbb{R}^{d_u}\right)$ as separable Banach spaces of function taking values in $\mathbb{R}^{d_a}$ and $\mathbb{R}^{d_u}$ respectively.\cite{beauzamy2011} The construction of a mapping, parameterized by  $\theta \in \Theta$, allows the Fourier neural operators to learn an approximation of $\mathcal{A} \rightarrow \mathcal{U}$. The optimal parameters $\theta^{\dagger} \in \Theta$ are determined through a data-driven empirical approximation.\cite{vapnik1999} The neural operators employ iterative architectures $v_0 \mapsto v_1 \mapsto \ldots \mapsto v_T$ where $v_j$ for $j=0,1, \ldots, T$ is a sequence of functions each taking values in $\mathbb{R}^{d_v}$.\cite{li2020b} The FNO architecture is shown in Fig.~\ref{NNFNO} which consists of three main steps.

(1) The input $a \in \mathcal{A}$ is lifted to a higher dimensional representation $v_0(x)=P(a(x))$ by the local transformation $P$ which is commonly parameterized by a shallow fully connected neural network.

(2) The higher dimensional representation $v_0(x)$ is updated iteratively by
	\begin{equation}
	v_{t+1}(x)=\sigma\left(W v_t(x)+\left(\mathcal{K}(a ; \phi) v_t\right)(x)\right), \quad \forall x \in D.	
	\label{eq9}
	\end{equation}
Where $\mathcal{K}: \mathcal{A} \times \Theta_{\mathcal{K}} \rightarrow \mathcal{L}\left(\mathcal{U}\left(D ; \mathbb{R}^{d_v}\right), \mathcal{U}\left(D ; \mathbb{R}^{d_v}\right)\right)$ maps to bounded linear operators on $\mathcal{U}\left(D ; \mathbb{R}^{d_v}\right)$ and is parameterized by $\phi \in \Theta_{\mathcal{K}}$, $W: \mathbb{R}^{d_v} \rightarrow \mathbb{R}^{d_v}$ is a linear transformation, and $\sigma: \mathbb{R} \rightarrow \mathbb{R}$ is non-linear activation function.

(3) The output $u \in \mathcal{U}$ is obtained by $u(x)=$ $Q\left(v_T(x)\right)$ where $Q: \mathbb{R}^{d_v} \rightarrow \mathbb{R}^{d_u}$ is the projection of $v_T$ and it is parameterized by a fully connected layer.\cite{li2020fourier} 

Denote $\mathcal{F}$ and $\mathcal{F}^{-1}$ as Fourier transform and its inverse transform of a function $f: D \rightarrow \mathbb{R}^{d_v}$ respectively. By substituting the kernel integral operator in Eq.~\ref{eq9} with a convolution operator defined in Fourier space, the Fourier integral operator can be rewritten as Eq.~\ref{eq10}. 
	\begin{equation}
	\left(\mathcal{K}(\phi) v_t\right)(x)=\mathcal{F}^{-1}\left(R_\phi \cdot\left(\mathcal{F} v_t\right)\right)(x), \quad \forall x \in D.	
	\label{eq10}
	\end{equation}
Where $R_\phi$ is the Fourier transform of a periodic function $\mathcal{K}: \bar{D} \rightarrow \mathbb{R}^{d_v \times d_v}$ parameterized by $\phi \in \Theta_{\mathcal{K}}$. The frequency mode $k \in \mathbb{Z}^d$. The finite-dimensional parameterization is obtained by truncating the Fourier series at a maximum number of modes $k_{\max }=\left|Z_{k_{\max }}\right|=\mid\left\{k \in \mathbb{Z}^d:\left|k_j\right| \leq k_{\max , j}\right.$, for $\left.j=1, \ldots, d\right\} \mid$. $\mathcal{F}\left(v_t\right) \in \mathbb{C}^{n \times d_v}$ can be obtained by discretizing domain $D$ with $n \in \mathbb{N}$ points, where $v_t \in \mathbb{R}^{n \times d_v}$.\cite{li2020fourier} By simply truncating the higher modes, $\mathcal{F}\left(v_t\right) \in \mathbb{C}^{k_{\max } \times d_v}$ can be obtained, here $\mathbb{C}$ is the complex space. $R_\phi$ is parameterized as complex-valued-tensor $({k_{\max } \times d_v \times d_v})$ containing a collection of truncated Fourier modes $R_\phi \in \mathbb{C}^{k_{\max } \times d_v \times d_v}$. Therefore, Eq.~\ref{eq11} can be derived by multiplying $R_\phi$ and $\mathcal{F}\left(v_t\right)$.
	\begin{equation}
	\left(R_\phi \cdot\left(\mathcal{F} v_t\right)\right)_{k, l}=\sum_{j=1}^{d_v} R_{\phi k, l, j}\left(\mathcal{F} v_t\right)_{k, j},\quad k=1, \ldots, k_{\max }, \quad j=1, \ldots, d_v.
	\label{eq11}
	\end{equation}


\subsection{U-net enhanced Fourier neural operator }
Wen et al.\cite{wen2022u} pointed out that FNO models may suffer from lower training accuracy due to the regularization impact of the FNO architecture in the multiphase flow problems.\cite{li2020fourier} They introduced an improved version of the Fourier neural operator, named U-FNO, which combines the strengths of both FNO-based and CNN-based models. The detailed description of the U-FNO network architecture is given in Appendix~\ref{app:NN}.

We propose a modified U-FNO architecture to better utilize the U-Net for learning small-scale flow structures, as shown in Fig.~\ref{NNUFNO}(b). The formulation of iterative network update is given by
	\begin{equation}
	v_{t+1}(x):=\sigma\left(W v_t(x)+\mathcal{F}^{-1}\left(R_\phi \cdot\left(\mathcal{F} v_t\right)\right)(x)+\mathcal{U^*} s_{t}(x)\right),\quad \forall x \in D .
	\label{eq14}
	\end{equation}
	\begin{equation}
	s_{t}(x) := v_{t}(x) - \mathcal{F}^{-1}\left(R_\phi \cdot\left(\mathcal{F} v_t\right) \right)(x),\quad \forall x \in D .
	\label{eq15}
	\end{equation}
Here, $s_{t}(x)\in\mathbb{R}^{d_v}$ denotes the small-scale flow field which can be obtained by subtracting the large-scale flow field from the original flow field $v(x)$. Then the U-Net $\mathcal{U^*}$ is used to learn the small-scale flow field. Finally, the full-field information transformed by $W$ is used to combine with FNO and U-Net, and then is connected with a nonlinear activation function $\sigma$ to form a new U-FNO network. 

Compared with the original U-FNO, our improved U-FNO performs better in 3D turbulence problems. Specifically, the minimum testing loss of original U-FNO and our modified U-FNO are 0.220 and 0.198, respectively. Therefore, the U-FNO mentioned later in this article refers to the modified U-FNO in Fig.~\ref{NNUFNO}(b).

\subsection{The implicit Fourier neural operator}
It has been demonstrated that with a large enough value of depth L, the FNO can serve as a universal approximator capable of accurately representing any continuous operator.\cite{kovachki2021} However, the increase of Fourier layers brings challenge for training the network due to the vanishing of gradient problem.\cite{you2022nonlocal} To overcome the shortage mentioned above, the idea of employing the shared hidden layer has been suggested.\cite{el2021implicit,winston2020monotone,bai2020multiscale} You et al.\cite{you2022learning} proposed the implicit Fourier neural operators (IFNOs) and demonstrated the technique of shallow-to-deep training. The detailed description of the IFNO network architecture is given in Appendix~\ref{app:NN}.

It should be noted that the numbers of parameters of the hidden layer are independent of the layers, which distinguishes it from the FNOs. It thus can greatly reduces the total number of parameters of the model and memory-usage. Furthermore, this architecture allows for the simple implementation of the shallow-to-deep initialization method.

With increasing depth of the layer $(1/L=\Delta t\rightarrow0)$, Eq.~\ref{eq13} can be regarded as a discrete version of ordinary differential equations (ODEs).\cite{you2022learning} Therefore, the network update can be reinterpreted as a discretization of a differential equation, and the optimal parameters obtained with $L$ layers can be served as the initial guess for deeper networks. The shallow-to-deep technique involves interpolating optimal parameters at depth $L$ and scaling them to maintain the final time of the differential equation.\cite{you2022learning} This technique can effectively improve the accuracy of the network and reduce the memory cost.  

\section{\label{sec:4}The implicit U-Net enhanced Fourier neural operator (IU-FNO)}
\begin{figure*}
	\includegraphics[width=1\linewidth]{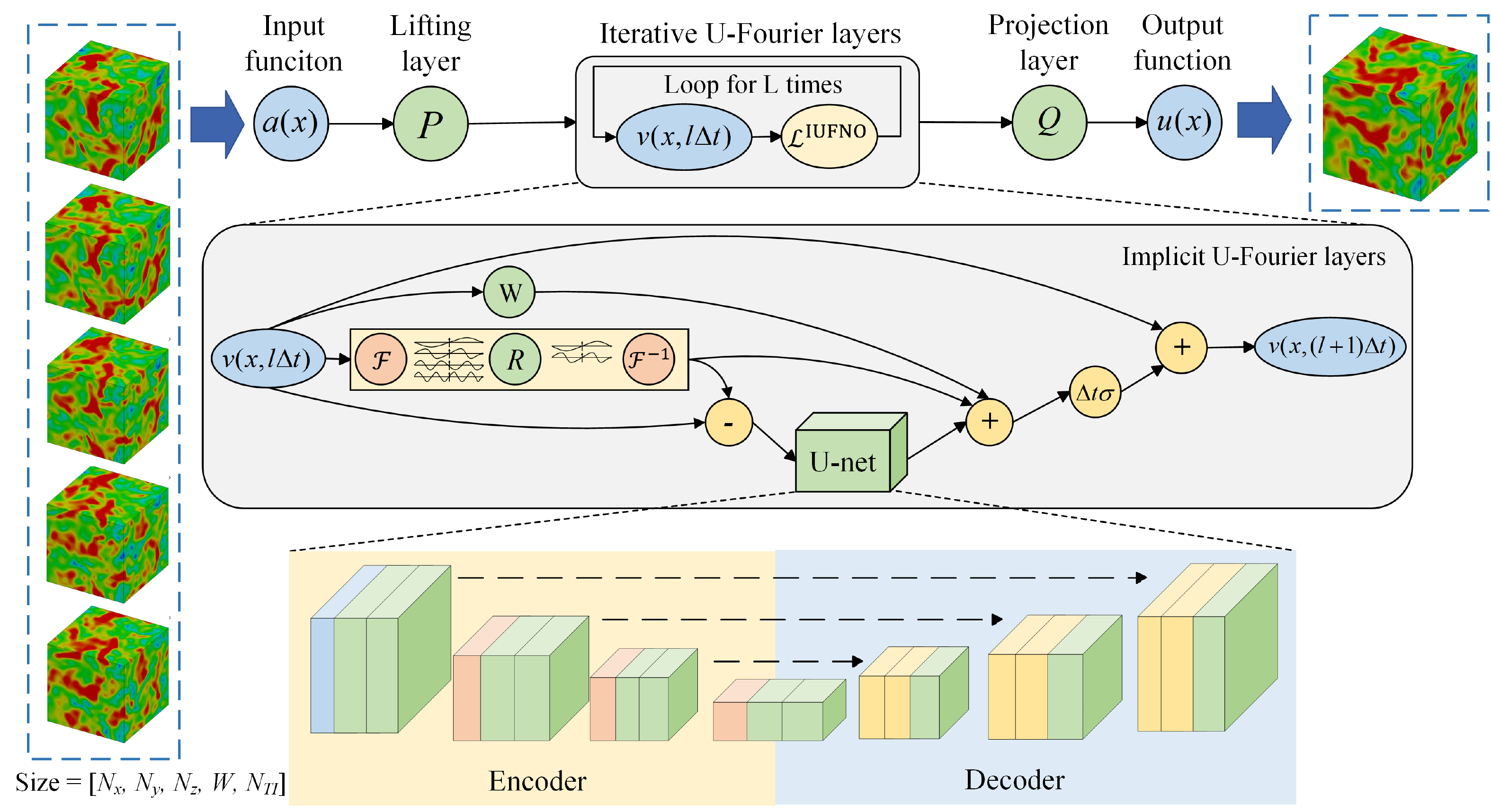}
	\caption{The architecture of implicit U-Net enhanced Fourier neural operator (IU-FNO).}
	\label{NNIUFNO}
\end{figure*}

We introduce an implicit U-Net enhanced Fourier neural operator (IU-FNO) to integrate the advantages of U-FNO and IFNO. The architecture of IU-FNO is shown in Fig.~\ref{NNIUFNO}. The velocity field from the first several time nodes is utilized as the input to the model, which is then converted into a high-dimensional representation via the lifting layer $P$. Then the velocity field is iteratively updated through the implicit U-Fourier layers, and finally the output is obtained through the projection of $Q$, which is the velocity field of the next time-node. The fundamental differences between the IU-FNO and FNO models are their network structures. FNO adopts a multilayer structure, where multiple Fourier layers with independent trainable parameters are connected in series. In contrast, the IU-FNO model utilizes a single Fourier layer with shared parameters and incorporates a U-net network to capture small-scale flow structures.

The formulation of iterative implicit U-Fourier layer update can be derived as
\begin{equation}
	\begin{aligned}
		{v}(x,(l+1) \Delta t) & =\mathcal{L}^{\rm IUFNO}[{v}({x}, l \Delta t)]:={v}({x}, l \Delta t)+\Delta t \sigma\left(c(x, l \Delta t) \right),\quad \forall x \in D,
	\end{aligned}
	\label{eq17}
\end{equation}
\begin{equation}
	c(x, l \Delta t) := W {v}({x}, l \Delta t)+ \mathcal{F}^{-1}\left(R_\phi \cdot\left(\mathcal{F} {v}({x}, l \Delta t)\right)\right)(x)+\mathcal{U^*} s(x, l \Delta t),\quad \forall x \in D,
	\label{eq18}
\end{equation}
\begin{equation}
	s(x,l \Delta t) := v(x,l \Delta t) - \mathcal{F}^{-1}\left(R_\phi \cdot\left(\mathcal{F} v(x,l \Delta t)\right) \right)(x),\quad \forall x \in D .
	\label{eq19}
\end{equation}
Here, $c(x,l \Delta t)\in\mathbb{R}^{d_v}$ has the global scale information of the flow field by combining large-scale information learned by FFT and small-scale information $s(x,l \Delta t)$ learned by the U-Net network $\mathcal{U}^*$. $s(x,l \Delta t)\in\mathbb{R}^{d_v}$ is obtained by subtracting the large-scale information from the complete field information $v(x,l \Delta t)$, shown in Eq.~\ref{eq19}. $\mathcal{U}^*$ is a CNN-based network, which provides a symmetrical structure with both an encoder and a decoder. The encoder is responsible for extracting feature representations from the input data, while the decoder generates the output signals.\cite{ronneberger2015u,wang2020towards} Furthermore, U-Net incorporates skip connections, enabling direct transmission of feature maps from the encoder to the decoder, thereby preserving the intricate details within the fields. The U-Net architecture has a relatively small number of parameters, such that its combination with FNO has a minimal effect on the overall numbers of parameters. Additionally, the implicit utilization of a shared hidden layer has significantly reduced the number of network parameters, which can make the network very deep.

\begin{table*}
	\caption{\label{parameters}Comparison of the numbers of parameters (calculated in Millions) with Fourier layer $L$ of different FNO-based models.}
	\begin{ruledtabular}
		\begin{tabular}{ccccc}
			Model& $L=4(T=4)$& $L=10$& $L=20$& $L=40$\\
			\hline
			FNO& 331.8M & 829.5M & 1659M & 3318M\\
			U-FNO& 332.0M & 830.0M & 1660M & 3320M\\
			IFNO& 82.97M & 82.97M & 82.97M & 82.97M\\ 
			IU-FNO& 83.02M & 83.02M & 83.02M & 83.02M\\ 
		\end{tabular}
	\end{ruledtabular}
\end{table*}
We compare the numbers of parameters with Fourier layer $L$ of different FNO-based models in Fourier mode equal to 20, as shown in Tab.~\ref{parameters}. The numbers of parameters of FNO and U-FNO models are 331.8 Million and 332.0 Million, respectively, when the number of Fourier layers is set to four. However, as the number of layers increases, the size of these parameters also increases, resulting in huge computational demands that can pose significant challenges for training. By using the implicit method of sharing hidden layers, the number of network parameters of the IFNO and IU-FNO models can be independent of the number of Fourier layers $L$. Specifically, the number of model parameters of IU-FNO is almost the same as that of IFNO. Moreover, it shows a significant reduction of approximately 75\% of parameter number compared with FNO.

\section{\label{sec:5}Numerical Examples}
In this section, the flow fields of the filtered direct numerical simulation (fDNS) of three types of turbulent flows are used for the evaluations of four FNO-based models, by comparing them against traditional LES with dynamic Smagorinsky model.\cite{yuan2023adjoint} The instantaneous snapshots of the fDNS data are employed for initializing the LES. The three types of turbulent flows includes forced homogeneous isotropic turbulence (HIT), temporally evolving turbulent mixing layer, and decaying HIT. 

In a \textit{posteriori} analysis, we perform the numerical simulations with ten different random initializations for each method in forced HIT, five different initializations in temporally evolving turbulent mixing layer, and five different initializations in decaying HIT respectively. We report the average value of the statistical results of different random initializations in the \textit{posteriori} analysis.

\subsection{\label{sec:5.1}Forced homogeneous isotropic turbulence}
The direct numerical simulation of forced homogeneous isotropic turbulence is performed with the uniform grid resolutions of $256^3$ in a cubic box of $(2\pi)^3$ with periodic boundary conditions.\cite{xie2020,yuan2020deconvolutional} The governing equations are spatially discretized using the pseudo-spectral method and a second-order two-step Adams-Bashforth explicit scheme is utilized for time integration.\cite{hussaini1987spectral,peyret2002spectral,chen1993statistical} The aliasing error caused by nonlinear advection terms is eliminated by truncating the high wavenumbers of Fourier modes by the two-thirds rule.\cite{hussaini1987spectral} The large-scale forcing is applied by fixing the velocity spectrum within the two lowest wavenumber shells in the velocity field to maintain the turbulence in the statistically steady state.\cite{yuan2020deconvolutional} The kinematic viscosity is adopted as $\nu=0.00625$, leading to the Taylor Reynolds number $Re_\lambda\approx100$. To ensure that the flow has reached a statistically steady state, we save the data after a long period (more than $10\tau$, here $\tau={L_I}/{u^{\rm r m s}}\approx1.0$ is large-eddy turnover times).

The DNS data is filtered into large-scale flow fields at grid resolutions of $32^3$ by the sharp spectral filter (described in Section~\ref{sec:2}) with cutoff wavenumber $k_c=10$. The time step is set to 0.001 and the snapshots of the numerical solution are taken every 200 steps as a time node. 45 distinct random fields are utilized as initial conditions, with 600 time nodes being saved for each group of computations. Therefore, the fDNS data with tensor size of $[45\times 600\times 32\times 32\times 32\times 3]$ can be obtained and serve as a training and testing dataset.\cite{li2022fourier} Specifically, the dataset we use to train the neural operator model consists of 45 groups, each group has 600 time nodes, and each time node denotes a filtered velocity field of $32^3$ with three directions.

Denotes the $m$-th time-node velocity field as $U_{m}$ and the $m$-th evolution increment filed as $\Delta U_m=U_{m+1}-U_m$ which is the difference of velocity field between two adjacent time nodes. The IU-FNO model takes the velocity fields of the previous five time nodes $[U_1, U_2, U_3, U_4, U_5]$ as input and produces the difference between the sixth and fifth velocity fields $[\Delta U_{5} = U_{6} - U_{5}]$ as output, as illustrated in Fig.~\ref{NNIUFNO}.\cite{li2022fourier} Once the predicted evolution increment $\Delta U_{5}^{\rm pre}$ is obtained from the trained model, the predicted sixth velocity field can be calculated by $U_{6}^{\rm pre} = U_{5} + \Delta U_{5}^{\rm pre}$. In the same way, $U_{7}^{\rm pre}$ can be predicted by $[U_2, U_3, U_4, U_5, U_{6}^{\rm pre}]$ and so on. Therefore, 600 time-nodes in each group can generate 595 input-output pairs ($[U_1, U_2, U_3, U_4, U_5\rightarrow U_{6}^{\rm pre}]$), and 45 groups can produce 26775 samples where we use 80\% for training and the rest of the samples for testing.\cite{li2022fourier}

All four data-driven models in this study utilize the same number of the Fourier modes, specifically a value of 20, and the initial learning rate is set to $10^{-3}$.\cite{peng2023linear} The Adam optimizer is used for optimization.\cite{kingma2014adam} The GELU function is chosen as the activation function.\cite{hendrycks2016gaussian} In order to ensure a fair comparison, the hyperparameters including learning rates and the decay rates are tuned for each method to minimize the training and testing loss, which is defined as
\begin{equation}
	\textit{Loss}=\frac{\|u^*-u\|_2}{\|u\|_2}, \text { where }\|\mathbf{A}\|_2=\frac{1}{n} \sqrt{\sum_{k=1}^n\left|\mathbf{A}_{\mathbf{k}}\right|^2}.
	\label{eqloss}
\end{equation}
Here, $u^*$ denotes the prediction of velocity fields and $u$ is the ground truth.

\begin{table*}
	\caption{\label{loss}Comparison of minimum training and testing loss with Fourier layer $L$ of different FNO-based models in forced homogeneous isotropic turbulence.}
	\begin{ruledtabular}
		\begin{tabular}{ccccc}
			\multicolumn{5}{c}{(Training Loss, Testing Loss)}\\
			\hline
			Model& $L=4(T=4)$& $L=10$& $L=20$& $L=40$\\
			\hline
			FNO& (0.225, 0.255) & N/A & N/A & N/A\\
			U-FNO& (0.174, 0.198) & N/A & N/A & N/A\\
			IFNO& (0.244, 0.261) & (0.216, 0.228) & (0.199, 0.214) & (0.185, 0.201)\\ 
			IU-FNO& (0.192, 0.211) & (0.171, 0.190) & (0.140, 0.163) & \textbf{(0.143, 0.155)}\\ 
		\end{tabular}
	\end{ruledtabular}
\end{table*}
A comparison of the minimum training and testing loss with Fourier layer $L$ of different FNO-based models in forced HIT is given in Tab.~\ref{loss}. It is shown that incorporating the U-net module to facilitate learning at small-scale information can improve the effectiveness of training and testing. Besides, the training and testing loss of the implicit method using the shared hidden layer (e.g. IFNO and IU-FNO) will be larger than FNO and U-FNO at layer $L=4$. However, as the number of hidden layer loop iterations $L$ increases, a significant reduction in the loss value is observed. The smallest testing loss value $0.155$ is obtained when $L$ equals 40 for IU-FNO. Therefore, the number of layer $L$ with minimum loss in each model is chosen for the \textit{posteriori} study. For FNO and U-FNO models, the Fourier layer number $L$ is set to 4, whereas IFNO and IU-FNO models have 40 implicit loop Fourier layers.

To avoid the over-fitting issue of the models, an additional independent ten groups of data from different initial fields are generated and utilized for the \textit{posteriori} evaluation. In the a \textit{posteriori} study, fDNS data is utilized as a baseline to evaluate various FNO-based models, including FNO, U-FNO, IFNO, and IU-FNO. The LES with DSM model is performed on the uniform grid with the grid resolution of $32^3$ in a cubic box of $(2\pi)^3$ using the same numerical method as DNS. The LES is initialized with the instantaneous velocity field obtained from fDNS. The DMM and VGM models adopt the same initial field and computational approach as the DSM model.

\subsubsection{The a \textit{posteriori} study}
\begin{figure*}
	\includegraphics[width=1\linewidth]{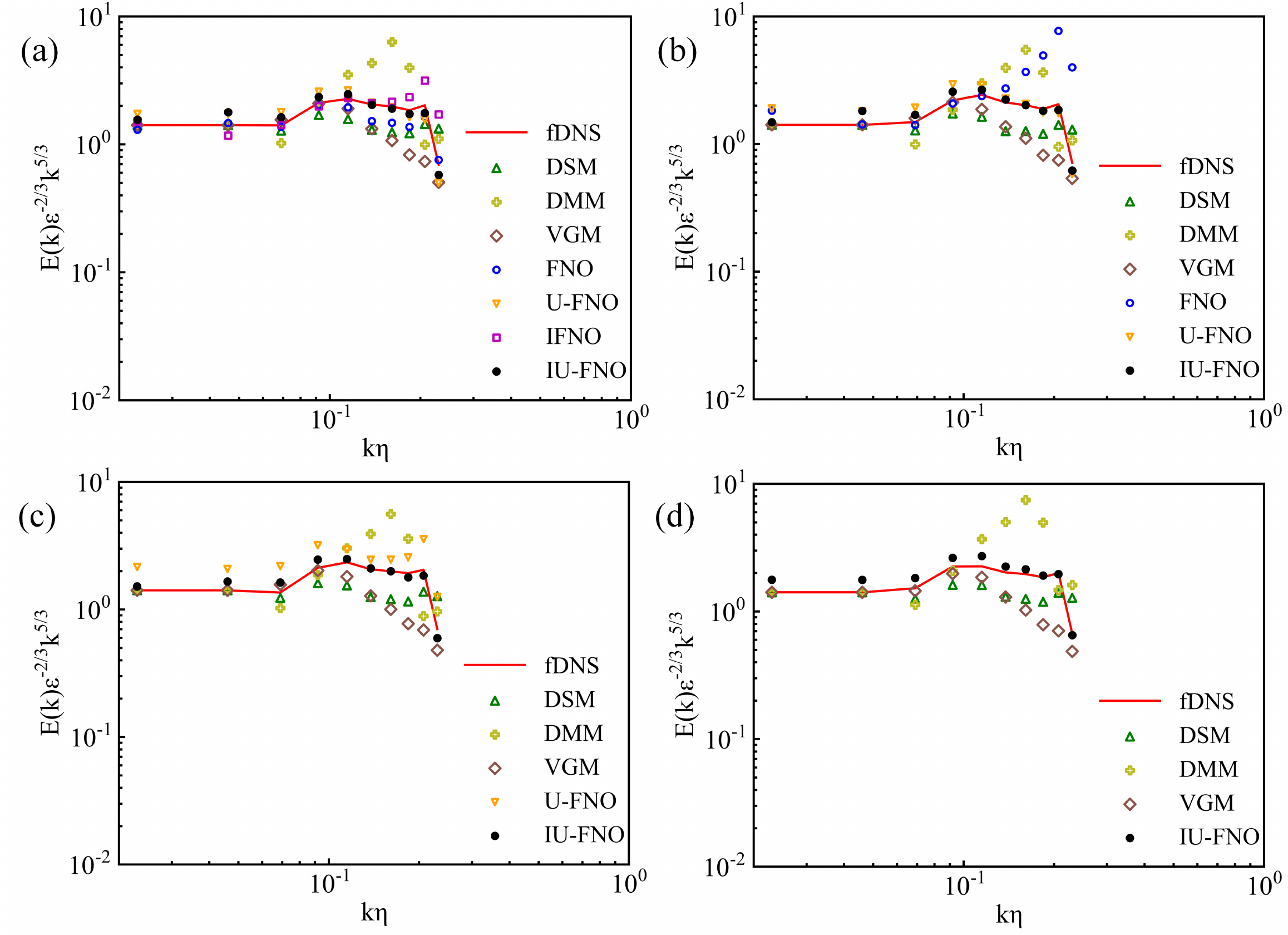}
	\caption{The normalized velocity spectra of LES using different models in the forced HIT at different time instants: (a)$t/\tau\approx4.0$; (b) $t/\tau\approx6.0$; (c)$t/\tau\approx8.0$; (d)$t/\tau\approx50.0$. Here, each prediction time instant for FNO-based model is $0.2\tau$.}
	\label{HIT_spectrum}
\end{figure*}
The normalized velocity spectra predicted by different FNO-based models and classical LES models at different time instants are shown in Fig.~\ref{HIT_spectrum}. Here, Kolmogorov length scale $\eta\approx0.023$ and the dissipation rate $\rm \varepsilon\approx0.825$ in the forced HIT are obtained from DNS data. For the traditional LES model DSM, the prediction errors become larger as the wavenumber $k$ increases. Specifically, the velocity spectrum at wavenumbers $4\le k \le 9$ is significantly lower than fDNS results. In terms of velocity spectrum prediction, the DMM model exhibits significantly higher values than the fDNS results in the wavenumbers $k$ range of 5 to 8. In contrast, the VGM model predicts much lower results for $k\geq4$. Overall, the normalized velocity spectrum predicted by the DSM model is more accurate than the DMM and VGM models. This study focuses on comparing the performance of FNO-based models, and we select the DSM model as the representative SGS model for comparison.

It can be seen from Fig.~\ref{HIT_spectrum}(a) that the normalized velocity spectrum predicted by data-driven models including FNO, U-FNO, and our proposed IU-FNO model are close to that of fDNS at time $t/\tau\approx4.0$. Here the large-eddy turnover time $\tau$ is provided in Eq.~(\ref{eq4}). However, the normalized velocity spectrum is overestimated by IFNO for the high wavenumbers at time $t/\tau\approx4.0$, and the prediction error becomes larger as the time increases. For the FNO and U-FNO models, large prediction errors have been identified at time $t/\tau\approx6.0$ and time $t/\tau\approx8.0$, respectively. It is worth noting that the prediction results of the FNO, IFNO, and U-FNO models are observed to be divergent with an increase in prediction time, and lose statistical significance. Therefore, we do not present these divergent results at the later time instants. On the contrary, IU-FNO always gives accurate predictions on the velocity spectrum in both short-term and long-term predictions for $t/\tau \le 50$. We also observe that IU-FNO is stable for $t/\tau \ge 100$ (not shown here).

\begin{figure*}
	\includegraphics[width=1\linewidth]{ 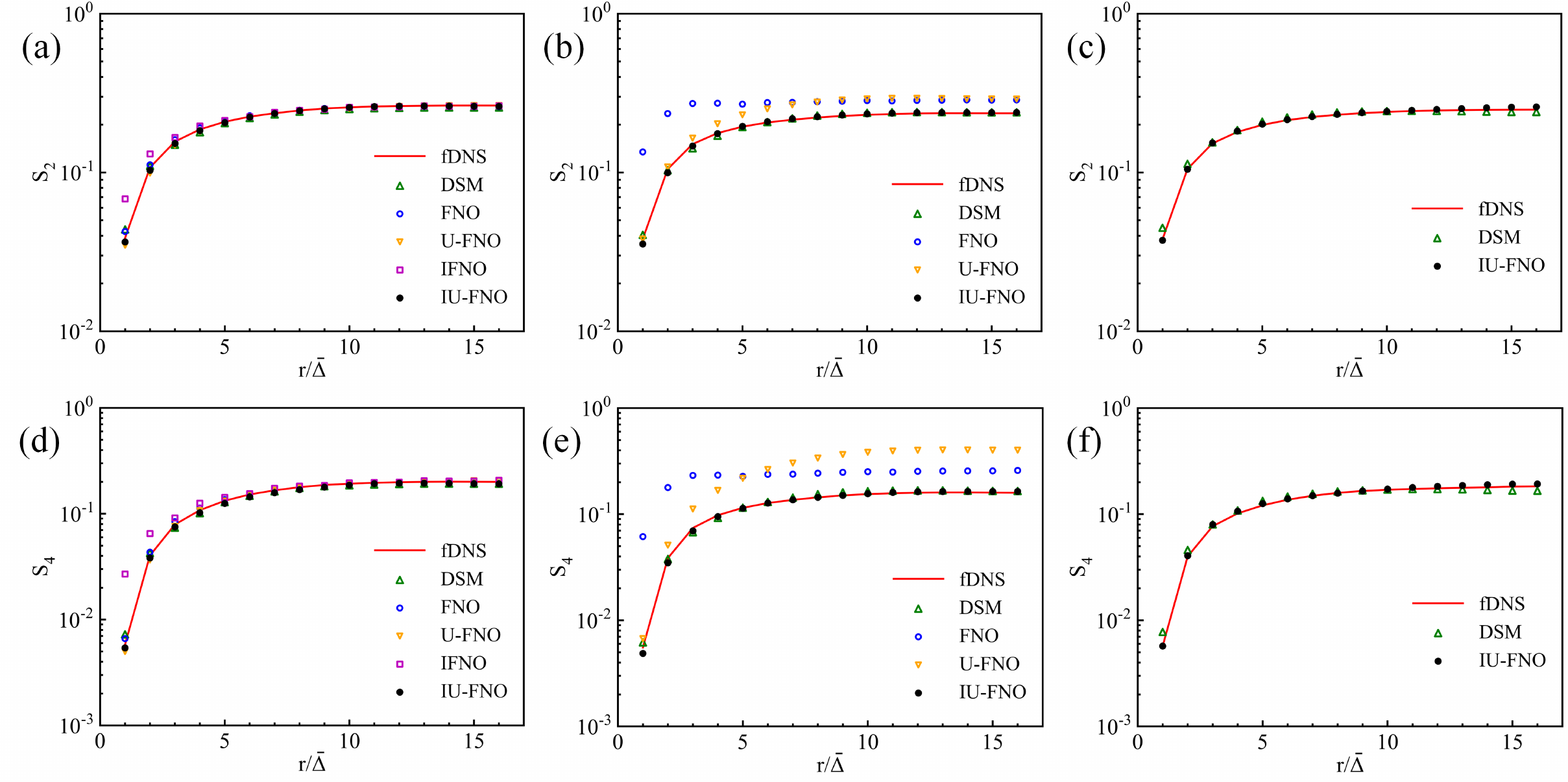}
	\caption{Second-order and fourth-order structure functions of the LES using different models in the forced HIT at different time instants: (a)Second-order, $t/\tau\approx4.0$; (b)Second-order, $t/\tau\approx8.0$; (c)Second-order, $t/\tau\approx50.0$; (d)Fourth-order, $t/\tau\approx4.0$; (e)Fourth-order, $t/\tau\approx8.0$; (f)Fourth-order, $t/\tau\approx50.0$. Here, each prediction time instant for FNO-based model is $0.2\tau$.}
	\label{HIT_structure}
\end{figure*}
To further examine the IU-FNO model in predicting multi-scale properties of turbulence, we compute the longitudinal structure functions of the filtered velocity, which are defined by \cite{xie2018modified,xie2020approximate}
\begin{equation}
	\bar{S}_n(r)=\left\langle\left|\frac{\delta_r \bar{u}}{\bar{u}^{\mathrm{rms}}}\right|^n\right\rangle,
	\label{eq20}
\end{equation}
where $n$ denotes the order of structure function and $\delta_r \bar{u}=[\overline{\mathbf{u}}(\mathbf{x}+\mathbf{r})-\overline{\mathbf{u}}(\mathbf{x})] \cdot \hat{\mathbf{r}}$ represents the longitudinal increment of the velocity at the separation $\mathbf{r}$. Here, $\hat{\mathbf{r}}=\mathbf{r} /|\mathbf{r}|$ is the unit vector. 

Fig.~\ref{HIT_structure} compares the second-order and fourth-order structure functions of the filtered velocity for different models with fDNS data at $t/\tau\approx4.0$, $t/\tau\approx8.0$, and $t/\tau\approx50.0$. It can be seen that the DSM model overestimates the structure functions at a small distances while underestimates them at large distances compared to those of the fDNS data. Moreover, as the time increases, the deviations of the structure functions predicted by the FNO, IFNO, and U-FNO models from the fDNS data become more serious. In contrast, the IU-FNO model can always accurately predict the structure functions at both small and large separations.

\begin{figure*}
	\includegraphics[width=1\linewidth]{ 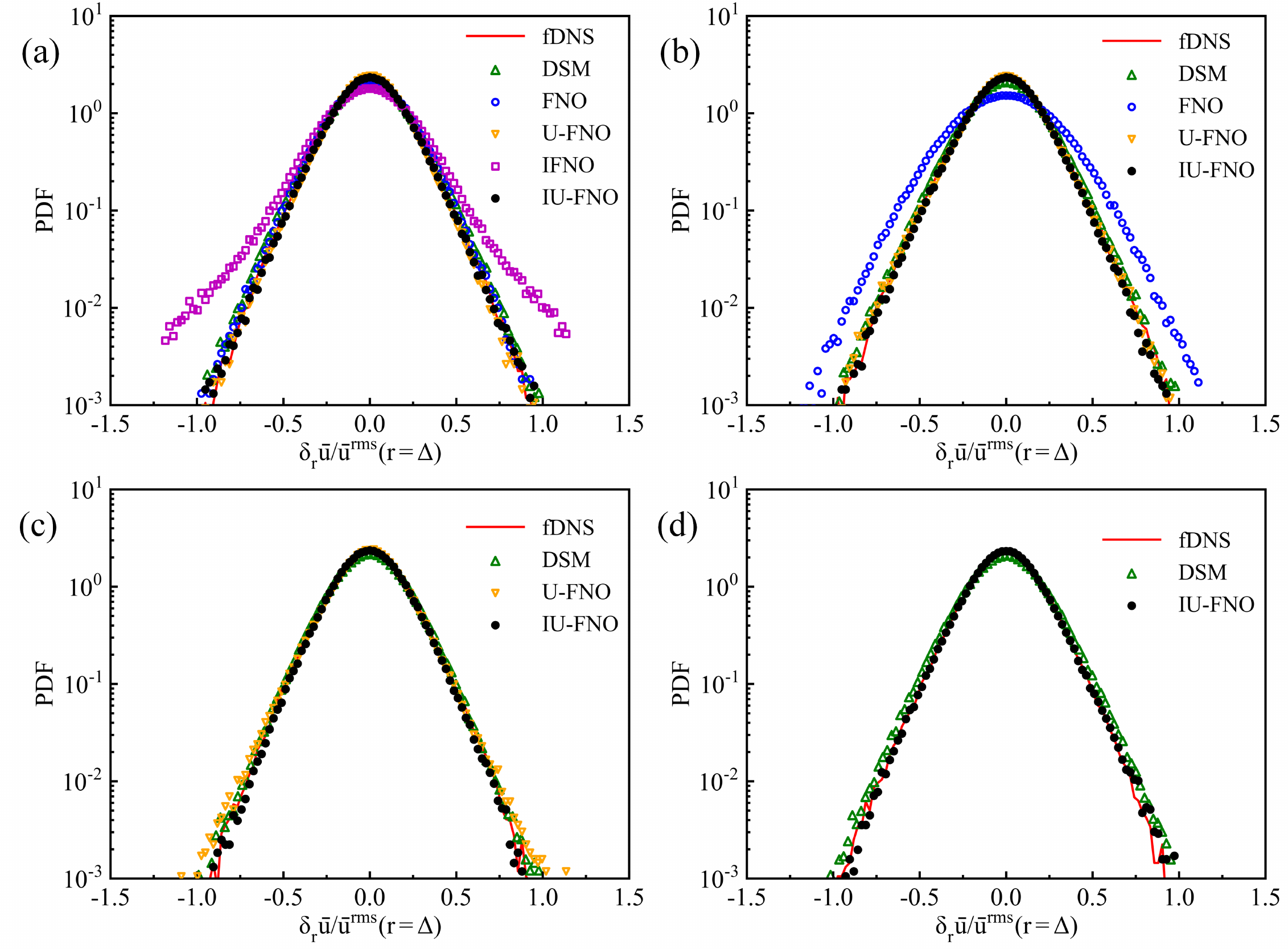}
	\caption{The PDFs of the normalized velocity increments $\delta_r \bar{u}/\bar{u}^{\rm rms} $ for LES using different models in the forced HIT at different time instants: (a)$t/\tau\approx4.0$; (b)$t/\tau\approx6.0$; (c)$t/\tau\approx8.0$; (d)$t/\tau\approx50.0$. Here, each prediction time instant for FNO-based model is $0.2\tau$.}
	\label{HIT_incv1}
\end{figure*}
Furthermore, we compare PDFs of the normalized velocity increments $\delta_r \bar{u}/\bar{u}^{\rm rms}$ with distance $r=\Delta$ at different time instants in Fig.~\ref{HIT_incv1}. It can be seen that the PDFs of the normalized velocity increments predicted by FNO, and U-FNO are in a good agreement with the fDNS data at the beginning, but the predicted PDFs become wider than the fDNS data as the time increases. The IFNO gives worst prediction on the PDF. The PDF predicted by the DSM model are also slightly wider than the fDNS results. The IU-FNO model gives the most accurate prediction on the velocity increments, demonstrating the excellent performance for both short-term and long-term predictions.

\begin{figure*}
	\includegraphics[width=1\linewidth]{ 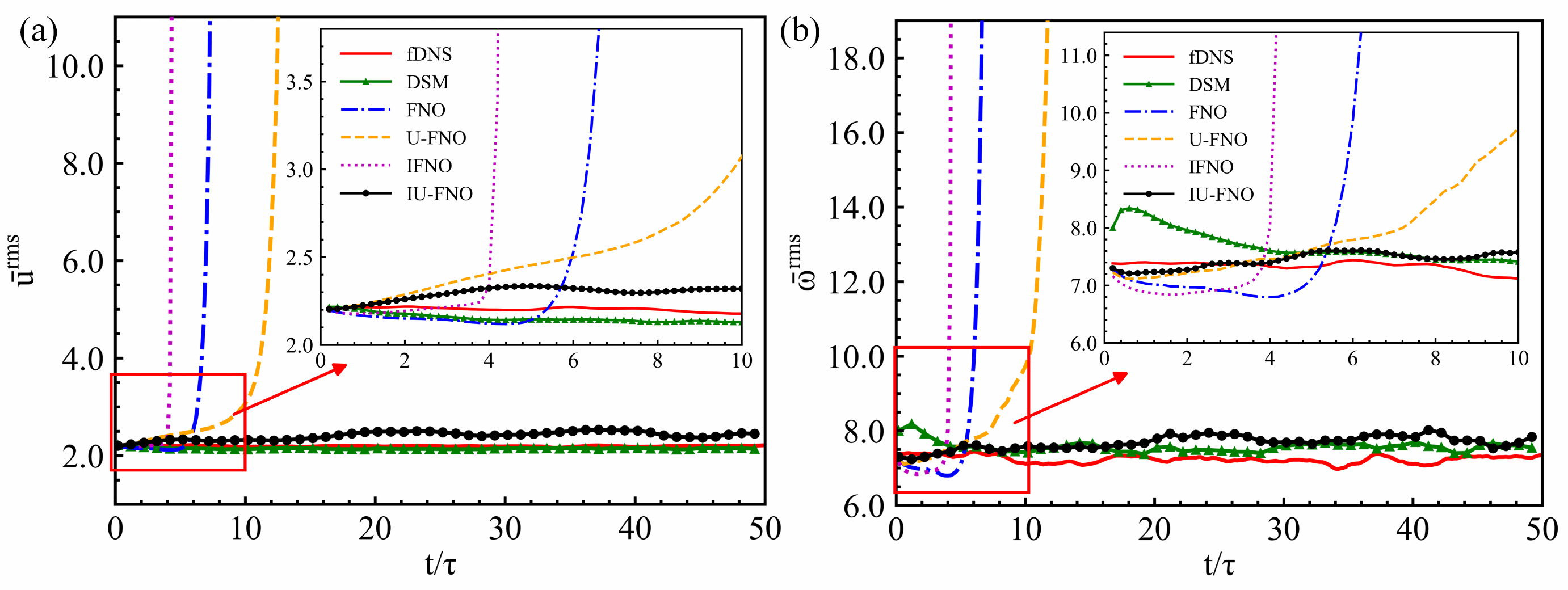}
	\caption{Temporal evolutions of the velocity rms value and vorticity rms value for LES using different models in the forced HIT. Here, each prediction time instant for FNO-based model is $0.2\tau$.}
	\label{HIT_uwRMS}
\end{figure*}
To further demonstrate the stability of different models, we display the evolution of the root-mean-square (rms) values of velocity and vorticity over time in Fig.~\ref{HIT_uwRMS}. Here, we plot the results from the $6$-th time instant. It can be seen that as the time increases gradually, the IFNO model will diverge quickly, followed by the FNO model, and the U-FNO model respectively. Since the predicted results will be used as the input for the next prediction, the prediction error will continue to be accumulated, which is one of the reasons why the data-driven model is difficult to be stable for a long-term prediction. The traditional DSM model is stable due to its dissipative characteristics.\cite{smagorinsky1963,kleissl2006numerical} Here, it is demonstrated that the proposed IU-FNO model can effectively and stably reconstrcut the long-term large-scale dynamics of the forced homogeneous isotropic turbulence.

\begin{figure*}
	\includegraphics[width=1\linewidth]{ 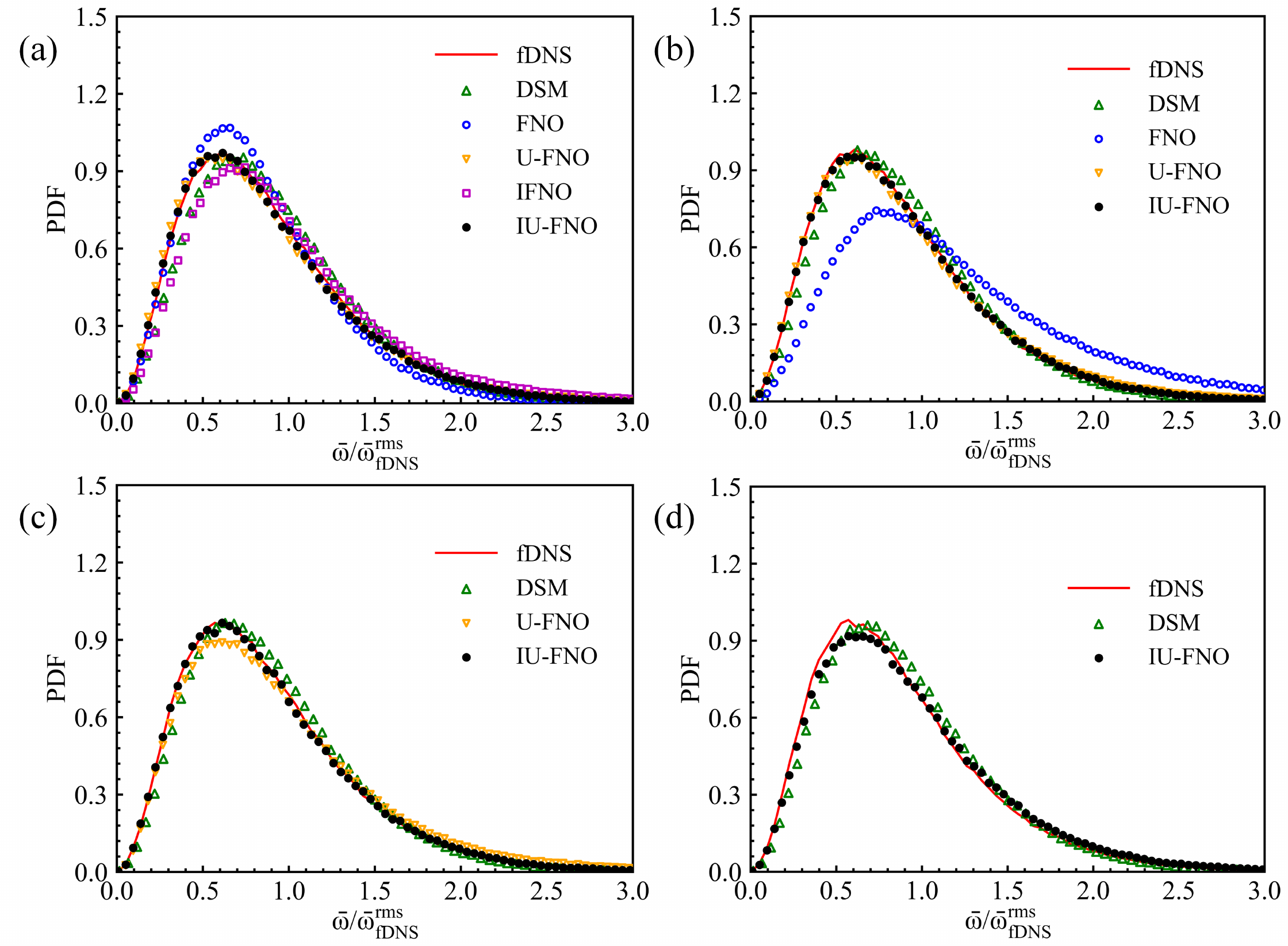}
	\caption{PDFs of the normalized vorticity $\bar{\omega}/\bar{\omega}_{\rm fDNS}^{\rm rms}$ for LES using different models in the forced HIT at different time instants:(a)$t/\tau\approx4.0$; (b)$t/\tau\approx6.0$; (c)$t/\tau\approx8.0$; (d)$t/\tau\approx50.0$. Here, each prediction time instant for FNO-based model is $0.2\tau$.}
	\label{HIT_vort}
\end{figure*}
The PDFs of the normalized vorticity magnitude at different time instants are shown in Fig.~\ref{HIT_vort}. Here, the vorticity is normalized by the rms values of the vorticity calculated by the fDNS data. It is shown that the PDFs predicted by all FNO-based models and DSM model have a reasonable agreement with those of fDNS in the short-term prediction for $t/\tau \le 4$. However, as the time increases, the deviation between the PDFs of $\bar{\omega}/\bar{\omega}_{\rm fDNS}^{\rm rms}$ predicted by FNO and U-FNO models and those of ground truth becomes more obvious. In contrast, the IU-FNO performs better than other FNO-based models in both short and long time predictions of vorticity statistics.

\begin{figure*}
	\includegraphics[width=1\linewidth]{ 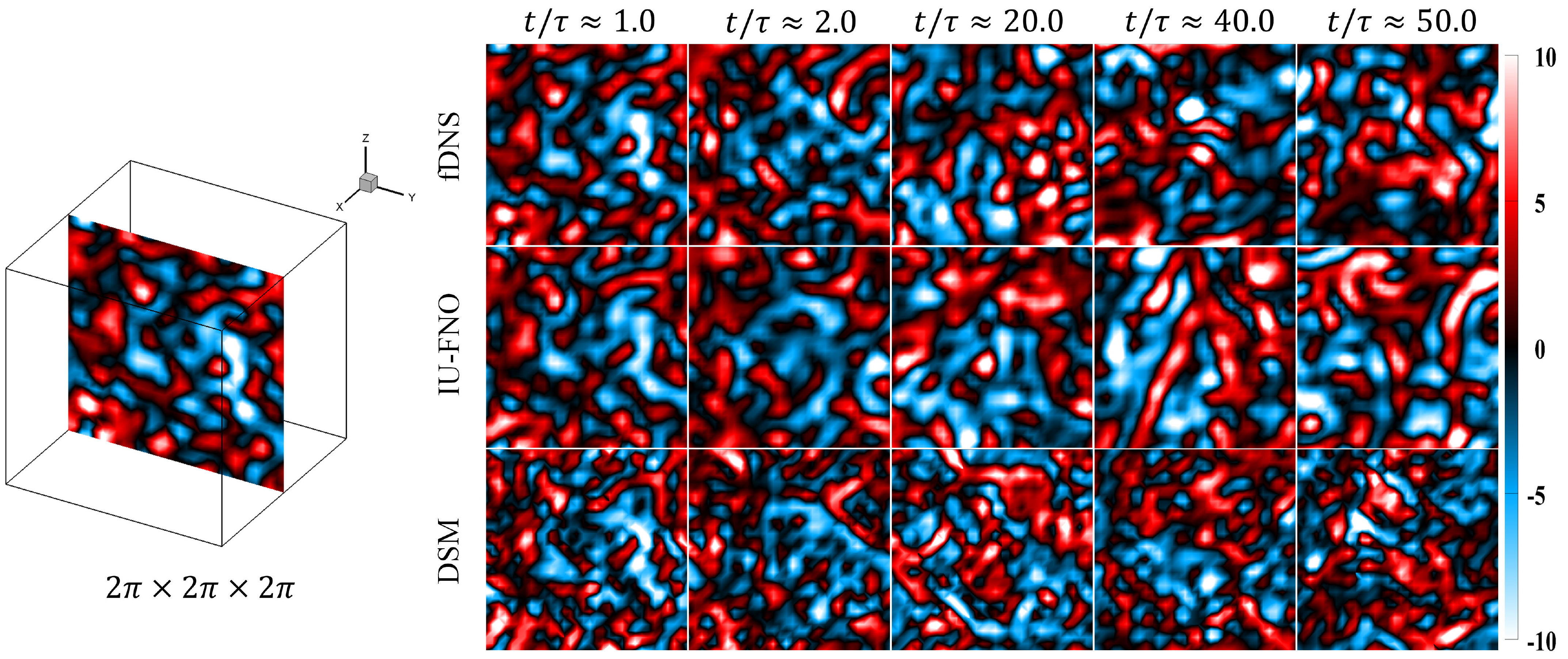}
	\caption{Evolution of predicted vorticity fields (at the center of y-z plane) as a function of time for forced HIT. Here, each prediction time instant for FNO-based model is $0.2\tau$.}
	\label{HIT_slice}
\end{figure*}
Fig.~\ref{HIT_slice} illustrates the contours of the vorticity fields predicted by different models. The instantaneous snapshots are selected on the center of the y-z plane at five different time instants. The DSM model produces factitious small-scale structures of vorticity, which significantly differ from those of the fDNS data. In contrast, vorticity fields given by IU-FNO model are very close to the benchmark fDNS results in the short-term prediction $t/\tau\approx1.0$ and $t/\tau\approx2.0$. Although the long-term prediction results of IU-FNO are not fully consistent with fDNS, its prediction of large-scale and small-scale structures is qualitatively better than that of the DSM model. Therefore, the IU-FNO model also performs better in long-term prediction of instantaneous vorticity structures, as compared to those of the DSM model.

\begin{figure*}
	\includegraphics[width=1\linewidth]{ 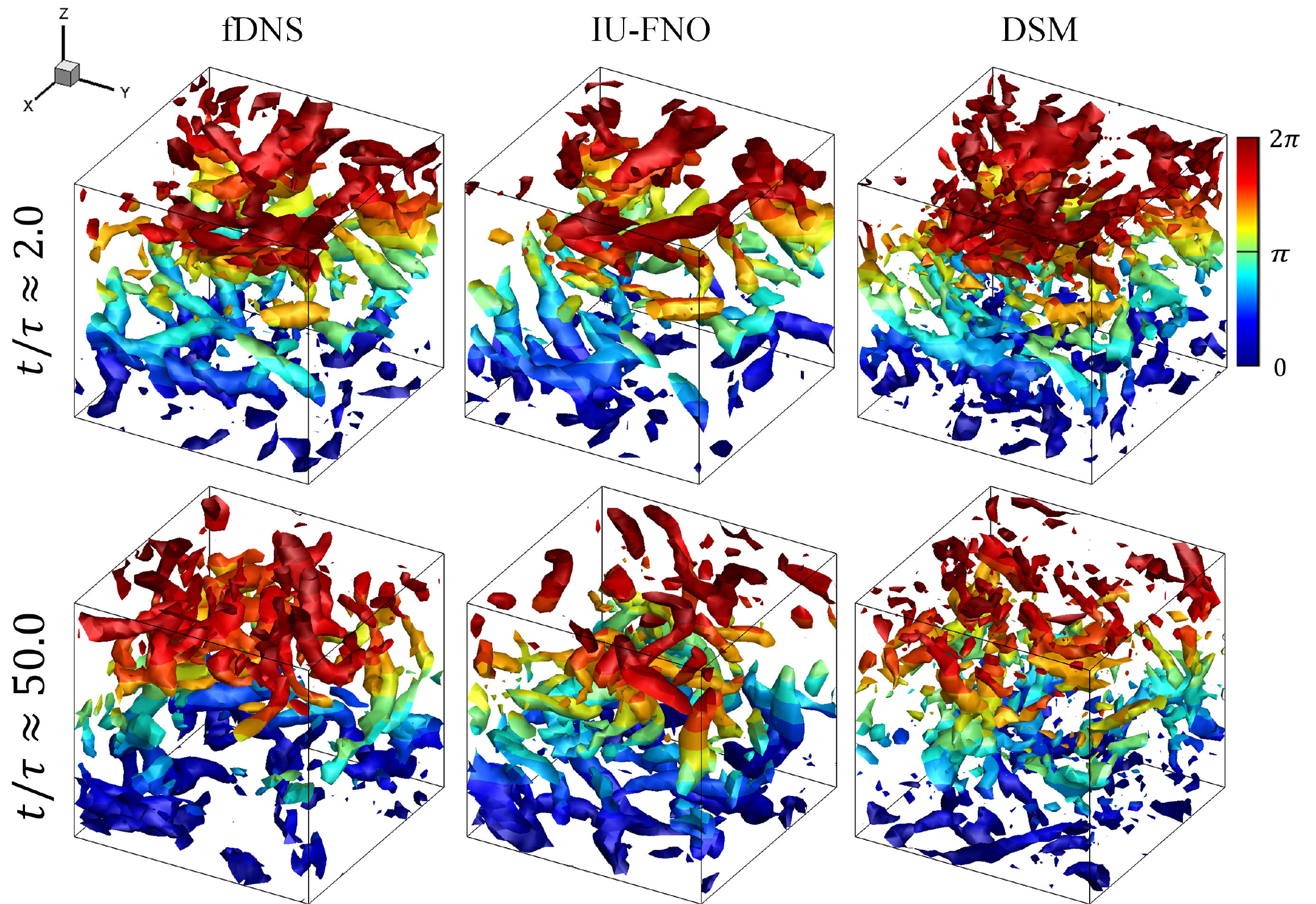}
	\caption{Isosurface of the normalized vorticity $\bar{\omega}/\bar{\omega}_{\rm fDNS}^{\rm rms}=1.5$(colored by altitude of z-direction) at time $t/\tau\approx2.0$ and $t/\tau\approx50.0$ for HIT. Here, each prediction time instant for FNO-based model is $0.2\tau$.}
	\label{HIT_wiso}
\end{figure*}
We demonstrate isosurfaces of the normalized vorticity $\bar{\omega}/\bar{\omega}_{\rm fDNS}^{\rm rms}=1.5$ colored by the altitude of z-direction in Fig.~\ref{HIT_wiso}. The spatial structures predicted by DSM and IU-FNO are compared to those of the fDNS data at time $t/\tau\approx2.0$ and $t/\tau\approx50.0$. It is revealed that the DSM model shows a limited accuracy in predicting the spatial structure of small-scale vortices. On the contrary, the IU-FNO model can better predict the overall flow structures of the vorticity field.

\subsubsection{Computational efficiency}
\begin{table*}
	\caption{\label{CE}Computational efficiency of different approaches on forced HIT.}
	\begin{ruledtabular}
		\begin{tabular}{ccccc}
			Method& Number of parameters(Million)& GPU memory-usage(MB)& GPU$\cdot$s& CPU$\cdot$s \\
			\hline
			DSM& N/A & N/A & N/A & 65.31 \\
			FNO& 331.8 & 3,204 & 0.058 & 2.953 \\
			U-FNO& 332.0 & 3,204 & 0.076 & 3.635 \\
			IFNO& 82.97 & 1,284 & 0.577 & 28.43 \\ 
			IU-FNO& 83.02 & 1,284 & 0.783 & 32.86 \\ 
		\end{tabular}
	\end{ruledtabular}
\end{table*}

Table.~\ref{CE} compares the computational cost of 10 prediction steps, number of parameters of the model, and GPU memory-usage for different FNO-based models on predictions of forced HIT. We carry out numerical simulations by using the Pytorch. The neural network models are trained and test on Nvidia Tesla V100 GPU, where the CPU type is Intel(R) Xeon(R) Gold 6240 CPU @2.60GHz. The DSM simulations are implemented on a computing cluster, where the type of CPU is Intel Xeon Gold 6148 with 16 cores each @2.40 GHz. Moreover, we conducted supplementary tests to measure the computational time for the FNO-based models using the same CPU as the DSM model. CPU$\cdot$s in Table.~\ref{CE} represents the time (second) required by each CPU core. Here, the FNO-based models are implemented on a single core CPU, while the DSM model is performed on CPU with 16 cores. So we assume that the CPU$\cdot$s of the DSM model is 16 times the actual time it takes. It can be seen that the FNO-based models are significantly more efficient than the DSM model. This is mainly attributed to the fact that the DSM model requires iterative solutions with a very small time step, while the FNO-based model can directly predict the flow state over a large time interval. In comparison to the original FNO, the IU-FNO model requires more computation time due to its deeper network. However, it is still about two times faster than the traditional DSM model. Table.~\ref{CE} also indicates that the computational efficiency of the FNO-based model can be further improved remarkably by using GPU. Moreover, the parameters and the GPU memory usage of the IU-FNO network model are reduced by about 75\% and 40\% compared with the original FNO model respectively. 

\subsubsection{Generalization on higher Taylor Reynolds numbers}
Here, we show that the IU-FNO model trained with low Taylor Reynolds number data can be directly used for the prediction of high Taylor Reynolds number cases without the need for additional training or modifications. The large-scale statistical features and flow structures in our simulations are observed to be insensitive to the Taylor Reynolds numbers.\cite{li2022fourier} We employ five sets of HIT data with different initial fields at a Taylor Reynolds number $Re_\lambda\approx250$. Owing to the large computational cost of DNS of turbulence at high Taylor Reynolds number, only 90 time nodes are computed for each initial field. Here, each time node for FNO-based model is $\rm \tau_h/3$, and the large-eddy turnover time is $\rm \tau_h \approx0.6$. For consistency, the computing devices for LES of DSM and IU-FNO are the same as those in the case of low Reynolds number. When performing the simulation for these higher Taylor Reynolds number cases on the same single CPU mentioned above, the DSM model required about 1900s to complete the task, while IU-FNO still only costs 32.26s. Thus, IU-FNO is more computationally efficient than DSM model, highlighting its considerable speed advantage. Here, DSM is performed on a higher grid resolution of $64^3$ to capture the small-scale fields at higher Taylor Reynolds number, but the result is still worse than the IU-FNO model.\cite{li2022fourier}

\begin{figure*}
	\includegraphics[width=1\linewidth]{ 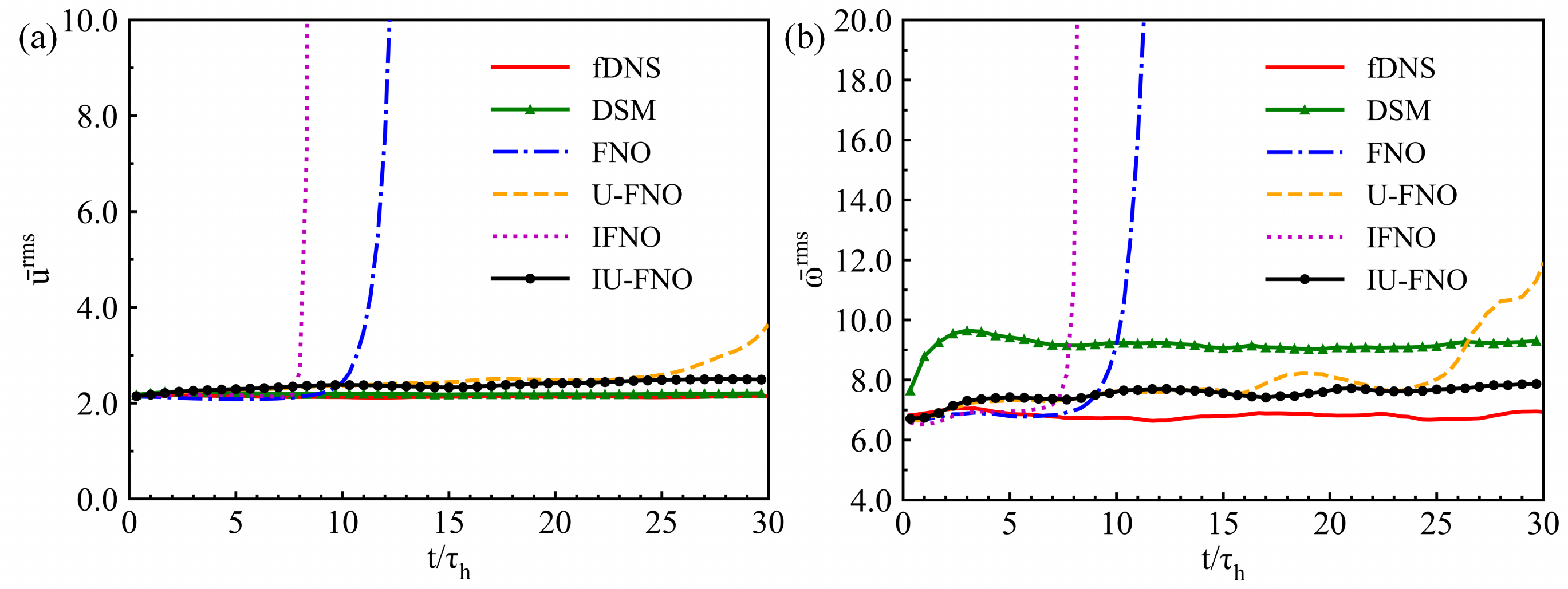}
	\caption{Temporal evolutions of the velocity RMS and vorticity RMS for LES using different models in the forced HIT at $Re_\lambda\approx250$.}
	\label{HIT_Re1q_uwRMS}
\end{figure*}
To assess the stability of the model at the high Reynolds number, we show the temporal evolutions of the rms values of velocity and vorticity in Fig.~\ref{HIT_Re1q_uwRMS}. It is revealed that the rms values of velocity and vorticity predicted by the IU-FNO and DSM models always perform stable, whereas other models become unstable as time increases. The rms values of velocity predicted by both the IU-FNO and DSM models show a good agreement with the fDNS data.

\begin{figure*}
	\includegraphics[width=1\linewidth]{ 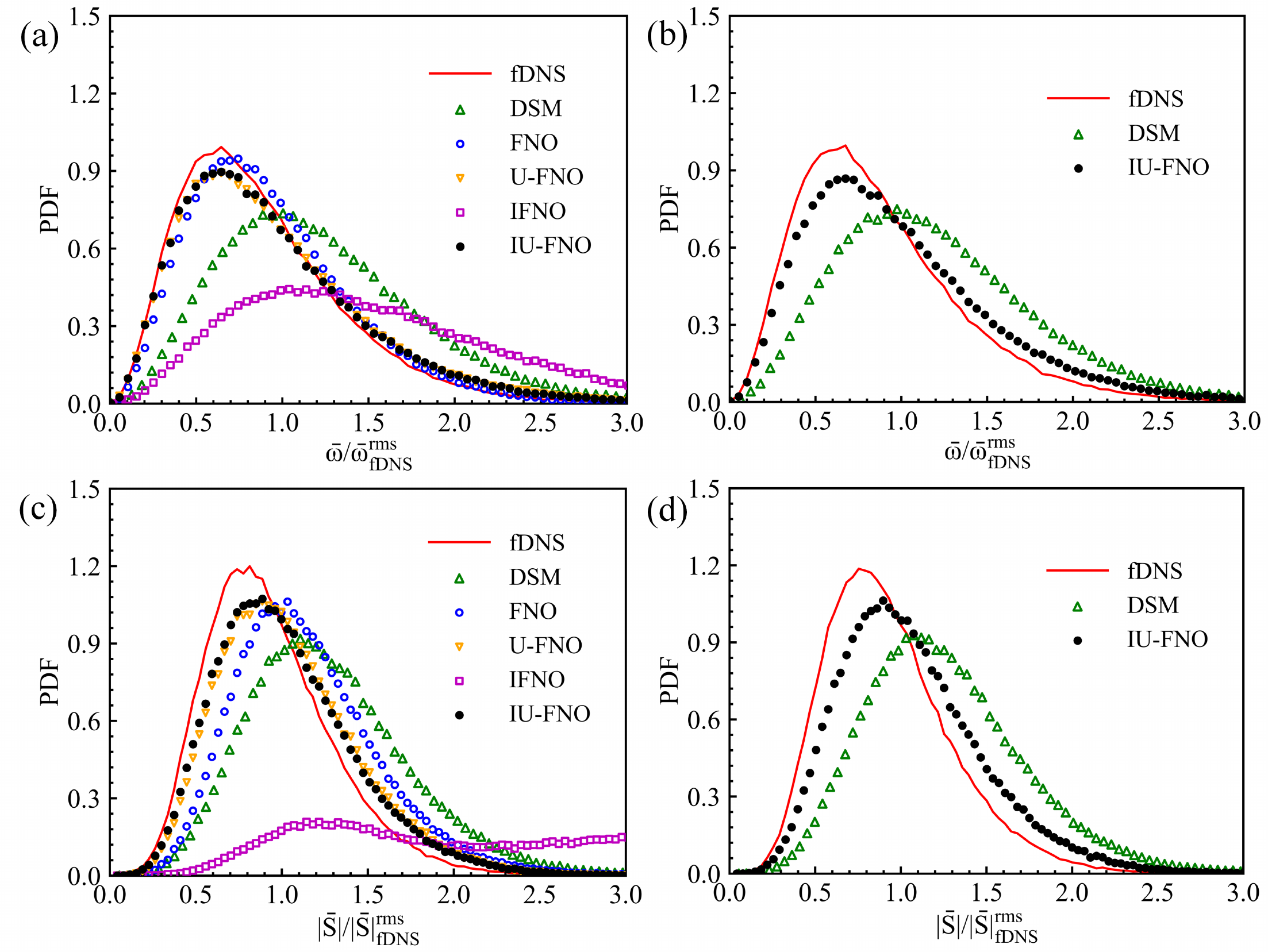}
	\caption{PDFs of the normalized vorticity $\bar{\omega}/\bar{\omega}_{\rm fDNS}^{\rm rms}$ and the normalized characteristic strain rate $|\rm \bar{S}|/|\rm \bar{S}|_{\rm fDNS}^{\rm rms}$ for LES using different models in the forced HIT at different time instants: (a)$\rm t/\tau_h\approx8.3$; (b)$\rm t/\tau_h\approx30$; (c)$\rm t/\tau_h\approx8.3$; (d)$\rm t/\tau_h\approx30$. Here, each prediction time instant for FNO-based model is $\rm \tau_h/3$.}
	\label{HIT_Re1q_vortst}
\end{figure*}
Furthermore, Fig~\ref{HIT_Re1q_vortst}(a) and (b) illustrate the PDFs of the normalized vorticity $\bar{\omega}/\bar{\omega}_{\rm fDNS}^{\rm rms}$. The IU-FNO model demonstrates a higher accuracy in predicting the peak and shape of PDFs than other models. The PDFs of the normalized characteristic strain rate of forced HIT at $t/\tau\approx8.3$ and $t/\tau\approx30$ are displayed in Fig.~\ref{HIT_Re1q_vortst}(c) and (d), respectively. Here, the  characteristic strain rate is defined by $|\bar{S}|=\sqrt{2 \bar{S}_{i j} \bar{S}_{i j}}$ and normalized by the rms values of the corresponding fDNS data. It is shown that the IU-FNO model outperforms the DSM model by accurately recovering both the peak value and overall shape of the PDF.

\subsection{Temporally evolving turbulent mixing layer}
In addition to benchmarking the performance of FNO-based models and classical SGS model on 3D forced HIT, we also evaluate their capabilities on a more complex simulation task: a 3D free-shear turbulent mixing layer. We focus on the comparison between the IU-FNO model and the conventional DSM model. The turbulent mixing layer provides a suitable example for studying the effects of non-uniform turbulent shear and mixing on subgrid-scale (SGS) models.\cite{wyp2022} 

The free-shear turbulent mixing layer is governed by the same Navier-Stokes equations (Eqs.~\ref{eq1} and \ref{eq2}) without the forcing term. The mixing layer is numerically simulated in a cuboid domain with lengths $L_1\times L_2\times L_3=8\pi\times 8\pi\times 4\pi$ using a uniform grid resolution of $N_1\times N_2\times N_3=256\times 256\times 128$. Here, $x_1 \in\left[-L_1 / 2, L_1 / 2\right]$, $x_2 \in\left[-L_2 / 2, L_2 / 2\right]$ and $x_3 \in\left[-L_3 / 2, L_3 / 2\right]$ denote the streamwise, normal, and spanwise directions, respectively. The initial streamwise velocity is given by\cite{wyp2022,wang2022compressibility,sharan2019turbulent}
\begin{equation}
	 u_1=\frac{\Delta U}{2}\left[\tanh \left(\frac{x_2}{2 \delta_\theta^0}\right)-\tanh \left(\frac{x_2+L_2 / 2}{2 \delta_\theta^0}\right)-\tanh \left(\frac{x_2-L_2 / 2}{2 \delta_\theta^0}\right)\right] + \lambda_1, 
	\label{eq21}
\end{equation}
where, $-L_2 /2 \leqslant x_2 \leqslant L_2 /2$, $\delta_\theta^0=0.08$ is the initial momentum thickness and $\Delta U = U_2 - U_1=2$ is the velocity difference between two equal and opposite free streams across the shear layer.\cite{wyp2022,yuan2023adjoint} The momentum thickness quantifies the range of turbulence region in the mixing layer, which is given by\cite{sharan2019turbulent,rogers1994direct,yuan2023adjoint}
\begin{equation}
	\delta_\theta=\int_{-L_2 / 4}^{L_2 / 4}\left[\frac{1}{4}-\left(\frac{\left\langle\bar{u}_1\right\rangle}{\Delta U}\right)^2\right] d x_2.
	\label{eqxita}
\end{equation}

The initial normal and spanwise velocities are given as $u_2=\lambda_2, u_3=\lambda_3$, respectively. Here, $\lambda_1, \lambda_2, \lambda_3 \sim$ $\mathcal{N}\left(\mu,\sigma^2\right)$, i.e., $\lambda_1, \lambda_2, \lambda_3$ satisfy the Gaussian random distribution. The expectation of the distribution is $\mu=0$ and the variance of the distribution is $\sigma^2=0.01$. The Reynolds number based on the momentum thickness $Re_\theta$ is defined as $Re_\theta=\Delta U \delta_\theta/{v_{\infty}}$. Here, the kinematic viscosity of shear layer is set to $v_{\infty}=5\times10^{-4}$, so the initial momentum thickness Reynolds number is $Re_{\theta}^0 = 320$.\cite{yuan2023adjoint} To mitigate the impact of the top and bottom boundaries on the central mixing layer, two numerical diffusion buffer zones are implemented to the vertical edges of the computational domain.\cite{wang2022compressibility,yuan2023adjoint,wyp2022} The periodic boundary conditions in all three directions are utilized and the pseudo-spectral method with the two-thirds dealiasing rule is employed for the spatial discretization. An explicit two-step Adam-Bashforth scheme is chosen as the time-advancing scheme.

The DNS data are then explicitly filtered by the commonly-used Gaussian filter, which is defined by\cite{pope2000,sagaut2006}
\begin{equation}
	G(\mathbf{r} ; \bar{\Delta})=\left(\frac{6}{\pi \bar{\Delta}^2}\right)^{1 / 2} \exp \left(-\frac{6 \mathbf{r}^2}{\bar{\Delta}^2}\right).
	\label{eq22}
\end{equation}
Here, the filter scale $\bar{\Delta}=8h_{\rm DNS}$ is selected for the free-shear turbulent mixing layer, where $h_{\rm DNS}$ is the grid spacing of DNS. The filter-to-grid ratio FGR=$\bar{\Delta}/h_{\rm LES}=2$ is utilized and then the corresponding grid resolution of fDNS: $64\times 64\times 32$ can be obtained.\cite{chang2022,wyp2022} The LES with DSM model is performed on the uniform grid with the grid resolution of $64\times 64\times 32$ in a cuboid domain with lengths $L_1\times L_2\times L_3=8\pi\times 8\pi\times 4\pi$ using the same numerical method as DNS and is initialized by the fDNS data.

We perform numerical simulations for 145 sets of distinct initial fields and save the results for 90 temporal snapshots for each initial field. The time interval for each snapshot is $200dt$, where $dt=0.002$ is the time step of DNS. Therefore, the data of size $[145\times 90\times 64\times 64\times 32 \times 3]$ can be obtained as training and testing sets. Similar to Section\ref{sec:5.1}, 80\% of data, including 9860 input-output pairs, is used for training, and 20\% of data, including 2465 pairs, is used for testing. 

\begin{figure}
	\includegraphics[width=1.0\columnwidth]{ 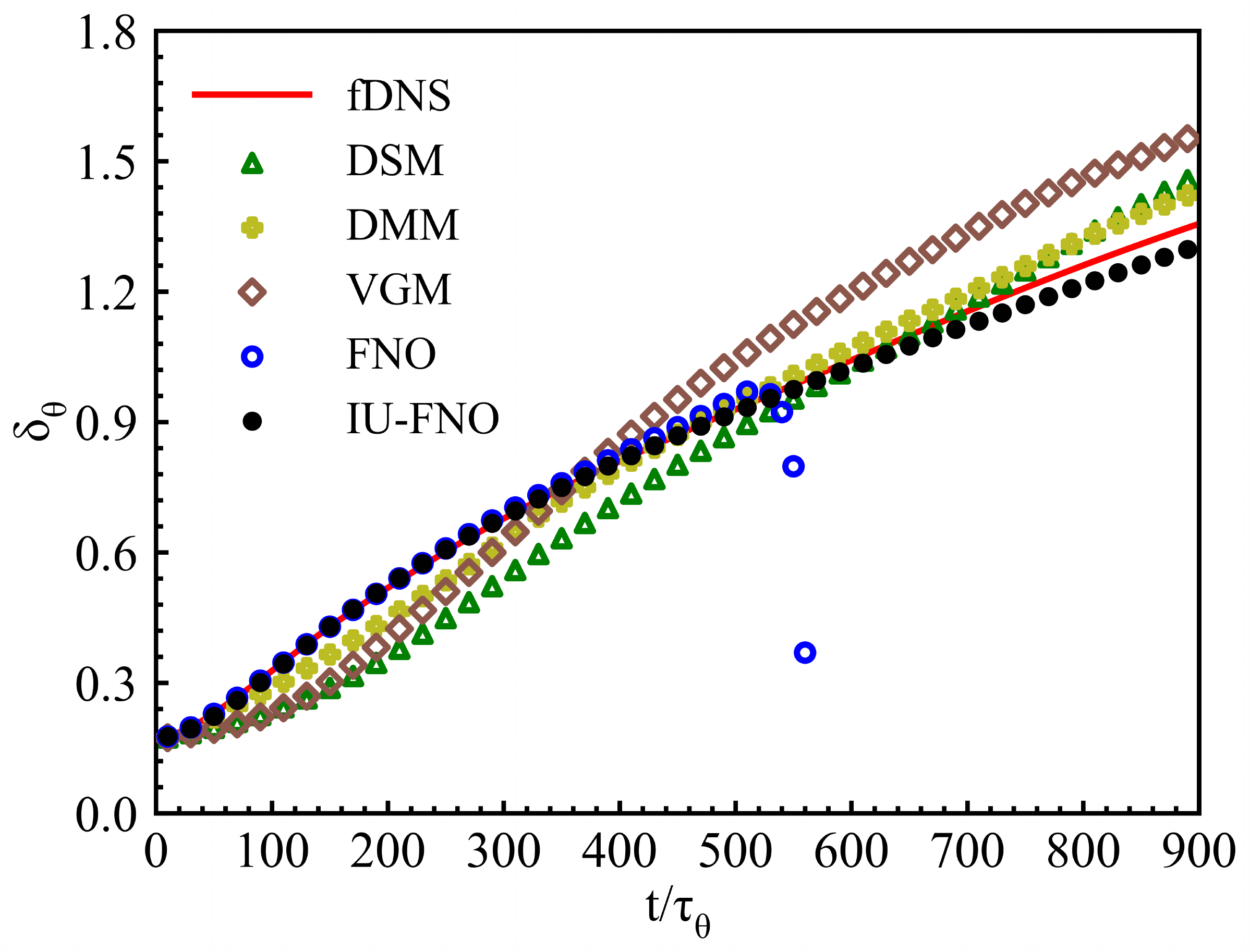}
	\caption{Temporal evolutions of the momentum thickness $\delta_\theta$ for LES using different models in the free-shear turbulent mixing layer.}
	\label{FS_theta}
\end{figure}
To perform a \textit{posteriori} analysis, we produce additional five sets of data with different initial fields, each containing ninety time nodes. Here, ninety time nodes are equivalent to nine hundred time units $(t/\tau_\theta=900)$ normalized by $\tau_\theta=\delta_\theta^0/\Delta U=20dt$. The temporal evolutions of the momentum thickness $\delta_\theta$ for LES using different models are shown in Fig.~\ref{FS_theta}. The DSM model underestimates the momentum thickness at the beginning of the transition region while overestimates the momentum thickness in the linear growth region. In comparison to the baseline fDNS, both the DMM and VGM models exhibit lower predicted values at the beginning, which gradually increase and eventually exceed the actual fDNS results. Since the three classical LES models exhibit similar performance, we select the DSM model for comparison with the FNO-based models. The FNO model shows a good ability to capture the momentum thickness growth rate during the early stages of temporal development. However, its prediction becomes invalid after 500 time units $(t/\tau_\theta \geqslant 500)$. In contrast, the predictions of the IU-FNO model always show a good agreement with fDNS in both transition and linear growth regions. 

\begin{figure*}
	\includegraphics[width=1\linewidth]{ 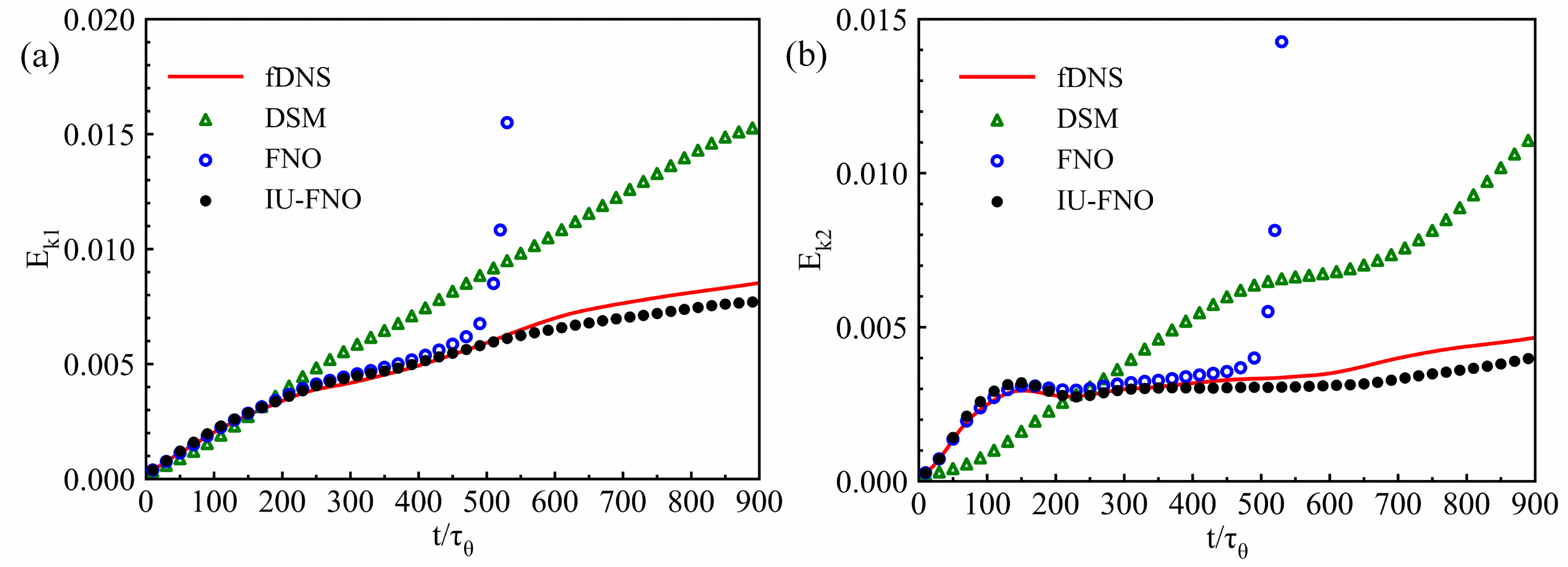}
	\caption{Temporal evolutions of the streamwise turbulent kinetic energy $E_{k1}$ and normal turbulent kinetic energy $E_{k2}$ for LES using different models in the free-shear turbulent mixing layer.}
	\label{FS_EK13}
\end{figure*}
Furthermore, the temporal evolutions of the streamwise turbulent kinetic energy $E_{k1}=\frac{1}{2}\left(\sqrt{\left\langle u_1 u_1\right\rangle}\right)^2$ and normal turbulent kinetic energy $E_{k2}=\frac{1}{2}\left(\sqrt{\left\langle u_2 u_2\right\rangle}\right)^2$ are displayed in Fig.~\ref{FS_EK13}. Here, $\langle\cdot\rangle$ denotes a spatial average over the whole computational domain. The turbulent kinetic energy in various directions increases gradually during the shear layer development in fDNS. Both streamwise and normal kinetic energy predicted by the DSM model are much larger than those of fDNS. FNO predicts reasonable results during the first 450 time units $(t/\tau_\theta\leq450)$, after that the results diverge quickly. By contrast, the IU-FNO model can well predict the kinetic energy in both streamwise and normal directions during the whole development of the shear layer, which is the closest to the fDNS data. 

\begin{figure*}
	\includegraphics[width=1\linewidth]{ 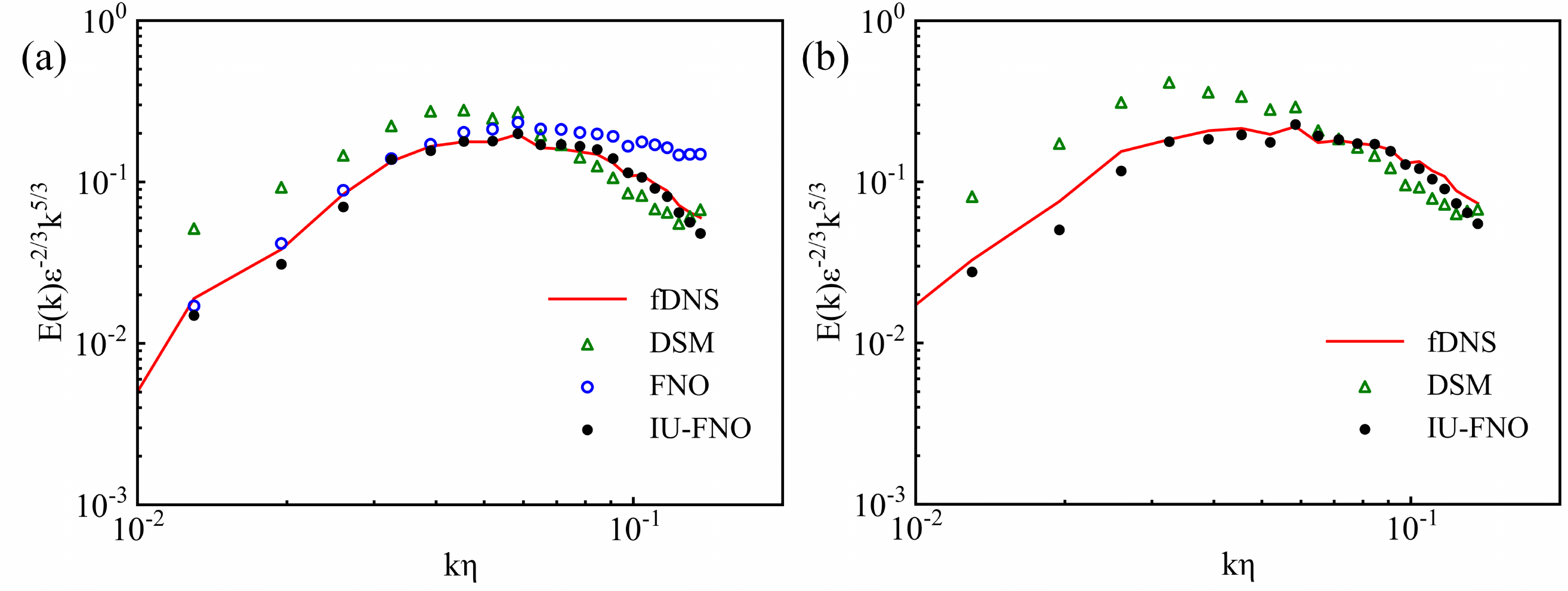}
	\caption{The normalized velocity spectra for LES using different models in the free-shear turbulent mixing layer at different time instants: (a)$t/\tau_\theta\approx 500$ (b)$t/\tau_\theta\approx 900$.}
	\label{FS_EK}
\end{figure*}
We then compare the normalized velocity spectrum of different models at time instants  $t/\tau_\theta\approx 500$ and $t/\tau_\theta\approx 900$, as shown in Fig.~\ref{FS_EK}. Here, Kolmogorov length scale $\eta\approx0.026$ at both $t/\tau_\theta\approx500$ and $t/\tau_\theta\approx900$. The dissipation rate $\rm \varepsilon\approx0.0023$ at $t/\tau_\theta\approx500$, and $\rm \varepsilon\approx0.0021$ at $t/\tau_\theta\approx900$ in the free-shear turbluent mixing layer. It can be seen that the normalized velocity spectrum predicted by the DSM model is overestimated at low wavenumbers and is underestimated at high wavenumbers when compared to those of the fDNS. The normalized velocity spectrum predicted by FNO is higher than benchmark fDNS at high wavenumbers, and the deviation will be larger as the time increases. In comparison, the IU-FNO model can accurately predict energy spectrum that agrees well with the fDNS data at various time instants.

\begin{figure*}
	\includegraphics[width=1\linewidth]{ 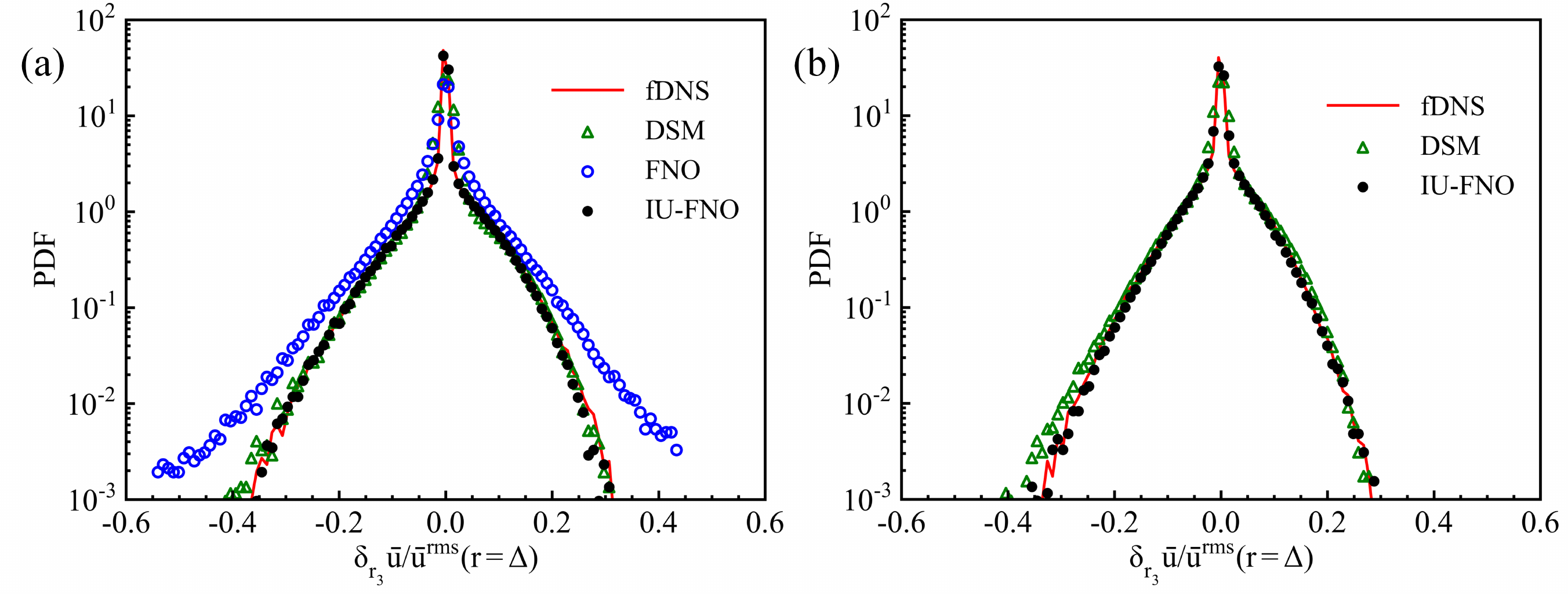}
	\caption{The PDFs of the spanwise velocity increment for LES using different models in the free-shear turbulent mixing layer at different time instants: (a)$t/\tau_\theta\approx 500$ (b)$t/\tau_\theta\approx 900$.}
	\label{FS_inc1}
\end{figure*}
Figure.~\ref{FS_inc1} illustrates the PDFs of velocity increment in the spanwise direction. Here, the spanwise velocity increment is given by $\delta_{r_3}\bar{u}=[\overline{\mathbf{u}}(\mathbf{x}+\mathbf{r})-\overline{\mathbf{u}}(\mathbf{x})] \cdot \hat{\mathbf{e}_3}$, where $\hat{\mathbf{e}_3}$ denotes the unit vector in the spanwise direction and the velocity increments are normalized by the rms values of velocity $\bar{u}^{\rm rms}$. The sharp peak of PDF is due to the non-turbulent regions where the velocity increment is nearly zero in the spanwise direction.\cite{wyp2022} The regions with non-zero velocity increments are predominantly governed by turbulence. It is shown that the IU-FNO model demonstrates better performance compared to both FNO and DSM models at different time instants.
	
\begin{figure*}
	\includegraphics[width=1\linewidth]{ 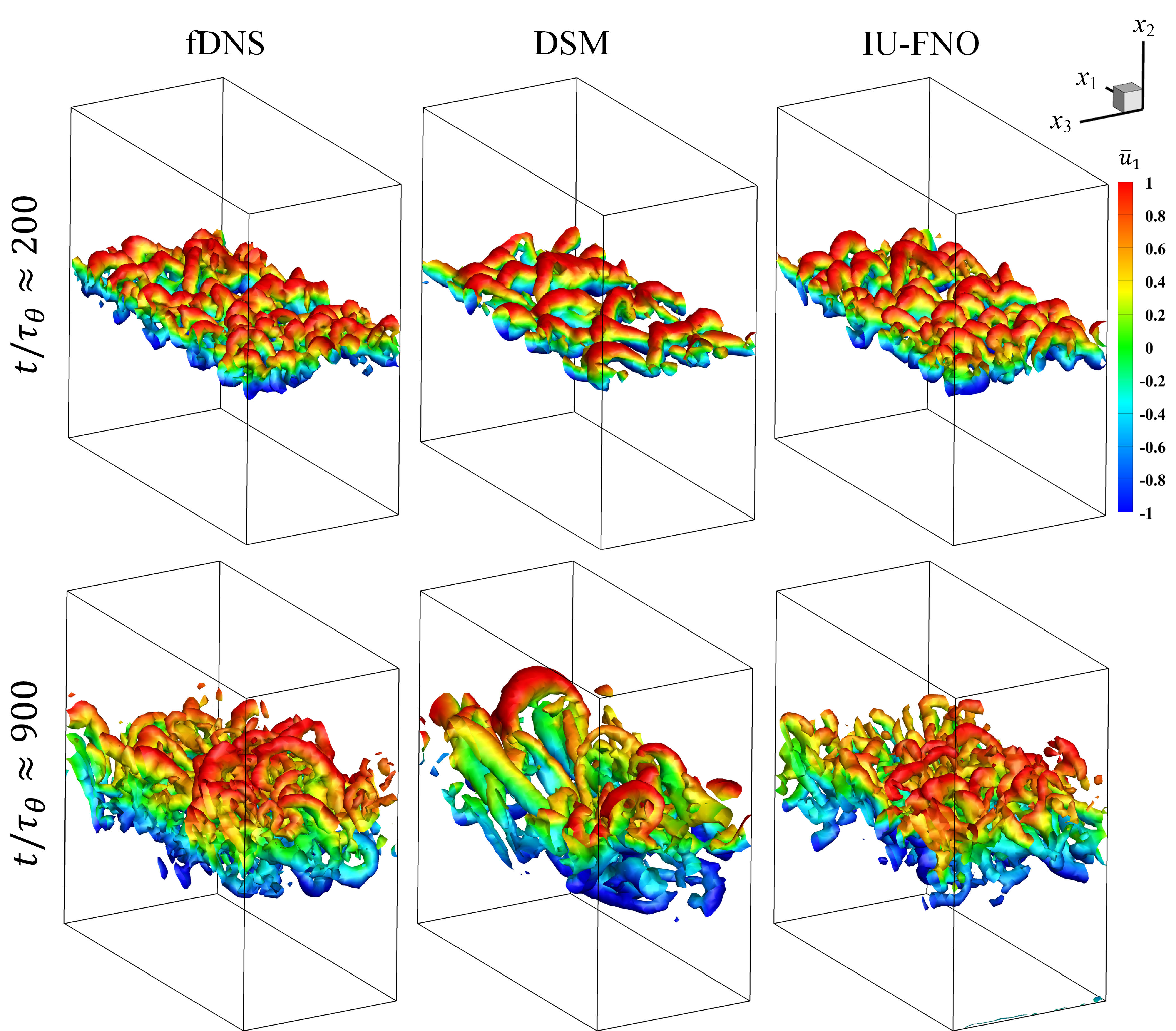}
	\caption{The iso-surface of the Q-criterion at $Q=0.2$ colored by the streamwise velocity at $t/\tau_\theta\approx 200$ and $t/\tau_\theta\approx 900$ in the free-shear turbulent mixing layer.}
	\label{FS_Q}
\end{figure*}
Finally, we compare the vortex structures predicted by the DSM model and IU-FNO model with fDNS data. The Q-criterion has been widely used for visualizing vortex structures in turbulent flows and is defined by\cite{hunt1988eddies,dubief2000coherent,zhan2019comparison}  
\begin{equation}
	Q=\frac{1}{2}\left(\bar{\Omega}_{i j} \bar{\Omega}_{i j}-\bar{S}_{i j} \bar{S}_{i j}\right),
	\label{eq23}
\end{equation}
where $\bar{\Omega}_{i j}=\left(\partial \bar{u}_i / \partial x_j-\partial \bar{u}_j / \partial x_i\right)/2$ is the filtered rotation-rate tensor. Fig.~\ref{FS_Q} displays the instantaneous isosurfaces of $Q=0.2$ at $t/\tau_\theta\approx 200$ and $t/\tau_\theta\approx 900$ colored by the streamwise velocity. It is observed that the DSM model predicts relatively larger vortex structures compared to the fDNS result. On the contrary, the IU-FNO model demonstrates a closer agreement with fDNS results especially in terms of reconstructing the small vortex structures, highlighting its advantage in improving the accuracy of LES.

\subsection{Decaying homogeneous isotropic turbulence}
We assess the extrapolation ability of different FNO-based models in LES of the decaying homogeneous isotropic turbulence (HIT). The numerical simulation of decaying HIT is conducted in a cubic box of $(2\pi)^3$ with periodic boundary conditions, and the numerical method is consistent with the forced HIT. The governing equations are spatially discretized using the pseudo-spectral method, incorporating the two-thirds dealiasing rule, at a uniform grid resolution of $N=256^3$. The temporal discretization scheme employs the explicit second-order two-step Adams-Bashforth method. We use the statistically steady flow field of the forced HIT as the initial field for the simulation of decaying turbulence. DNS of decaying turbulence is performed over about six large-eddy turnover times$(\tau=L_I/u^{\rm rms})$. In order to assess the extrapolation ability of different FNO-based models, only the flow fields at first two large-eddy turnover times $t/\tau \leq 2$ are used for training, and the flow fields at $t/\tau > 2$ are in the unseen flow regime where the magnitude of velocity fluctuation is different from the training data. 

The kinematic viscosity is set to $\nu=0.00625$ and the initial Taylor Reynolds number is $Re_\lambda\approx 100$. The sharp spectral filter(mentioned in Section \ref{sec:2}) with cutoff wavenumber $k_c=10$ is used to filter the DNS data. Here, we calculated 595 different sets of initial fields and stored a snapshot every $0.1\tau$. Finally, the fDNS data of size $[595\times 20\times 32\times 32\times 32\times 3]$ can be obtained, and 80\% of datas are used for training and 20\% for testing.

\begin{figure*}
	\includegraphics[width=1\linewidth]{ 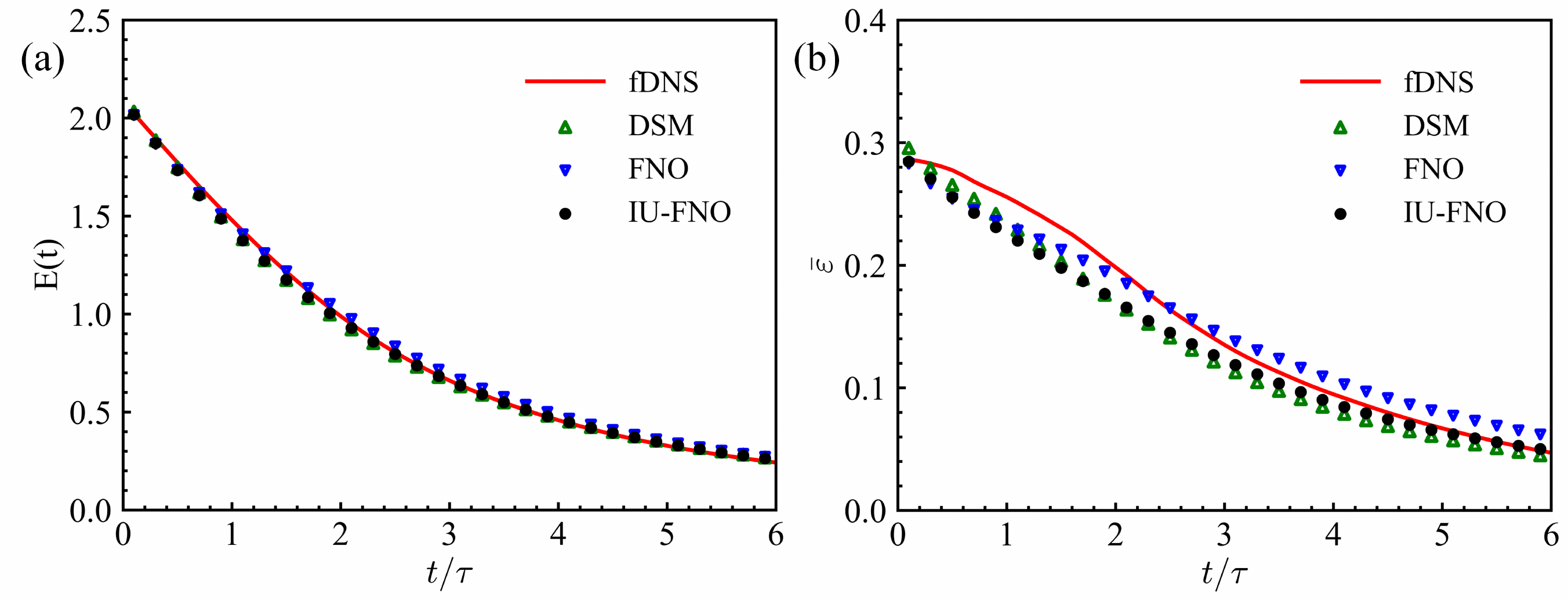}
	\caption{Temporal evolutions of the turbulent kinetic energy $E(t)$ and the average dissipation rate $\bar{\varepsilon}$ for different models in decaying HIT.}
	\label{DC_Ekep}
\end{figure*}
After training, five more groups of data with different initial fields are generated to perform a \textit{posteriori} analysis. Figure.~\ref{DC_Ekep} compares the temporal evolutions of the turbulent kinetic energy $E(t)=\int_0^{\infty}E(k)dk=\frac{1}{2}\left(u^{\rm rms}\right)^2$ and the resolved dissipation rate $\bar{\varepsilon}$ of DSM, FNO, and IU-FNO models with fDNS data. Here, the dissipation rate is defined by $\bar{\varepsilon}=2\nu\left\langle\bar{S}_{i j} \bar{S}_{i j}\right\rangle$. It can be seen that the kinetic energy gradually decays from the initial state over time, and all models can predict the turbulent kinetic energy well in the short period. However, the dissipation rate predicted by IU-FNO is more accurate than those of the DSM and FNO models at $t/\tau\geqslant4$. 

\begin{figure*}
	\includegraphics[width=1\linewidth]{ 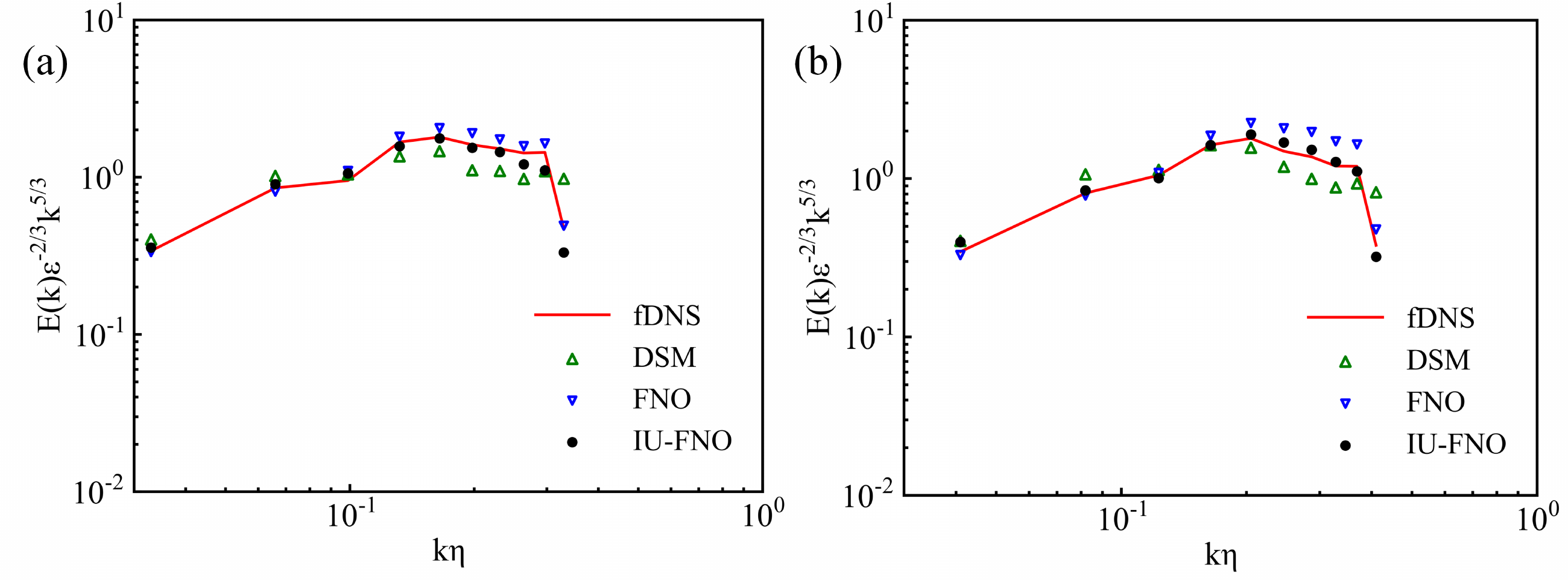}
	\caption{The normalized velocity spectra for different models in decaying HIT at $t/\tau\approx4.0$ and $t/\tau\approx6.0$.}
	\label{DC_spec}
\end{figure*}

Further, we evaluate the normalized velocity spectra for different models at two different time instants $t/\tau\approx4.0$ and $t/\tau\approx6.0$ in Fig.~\ref{DC_spec}. Here, Kolmogorov length scale $\eta\approx0.033$ at $t/\tau\approx4.0$, and $\eta\approx0.041$ at $t/\tau\approx6.0$. The dissipation rate $\rm \varepsilon\approx0.214$ at $t/\tau\approx4.0$, and $\rm \varepsilon\approx0.085$ at $t/\tau\approx6.0$ in the decaying HIT. The kinetic energy at all wavenumbers decreases with the time. The DSM model overestimates the kinetic energy at low wavenumbers. The FNO model overpredicts the energy spectrum at all wavenumbers. In contrast, the normalized velocity spectrum predicted by the IU-FNO model is in a good agreement with fDNS data.

\begin{figure*}
	\includegraphics[width=1\linewidth]{ 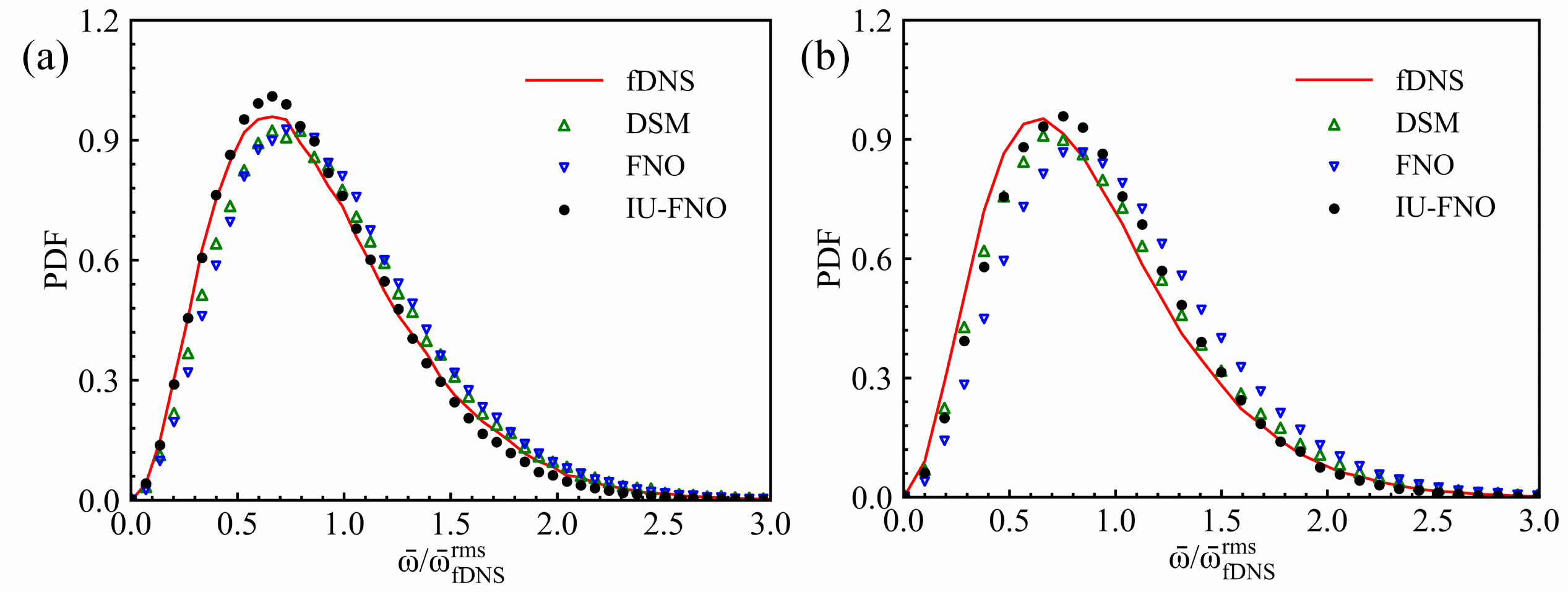}
	\caption{PDFs of the normalized vorticity $\bar{\omega}/\bar{\omega}_{\rm fDNS}^{\rm rms}$ for different models in decaying HIT at $t/\tau\approx4.0$ and $t/\tau\approx6.0$.}
	\label{DC_vort}
\end{figure*}
Finally, the PDFs of the normalized vorticity at the dimensionless time $t/\tau\approx4.0$ and $t/\tau\approx6.0$ are depicted in Fig.~\ref{DC_vort}. The rms values of the vorticity calculated by the fDNS data are used for normalization. Both the DSM model and FNO model give the wrong prediction of the peak location. On the contrary, the IU-FNO model slightly outperforms these models at both $t/\tau\approx4.0$ and $t/\tau\approx6.0$, which provides a reasonably accurate prediction for both the locations and peaks of the PDFs of the vorticity.

\section{\label{sec:6}Discussion and future work}
Simulations of three-dimensional (3D) nonlinear partial differential equations (PDEs) are of great importance in engineering applications. While data-driven approaches have been widely successful in solving one-dimensional (1D) and two-dimensional (2D) PDEs, the relevant works on data-driven fast simulations of 3D PDFs are relatively rare. The need for significant model complexity and a large number of parameters to accurately model the non-linear interactions in 3D PDEs (including turbulent flows) is a major challenge. In such situations, training and implementing neural networks may not be as efficient as traditional numerical methods. 

Recently, the FNO has proven to be a highly effective surrogate model in solving PDEs, indicating its significant potential for addressing 3D nonlinear problems.\cite{li2020fourier,li2022fouriergeo,de2022cost} The utilization of FNO in 3D turbulence has attracted more and more attention. Li et al. utilized FNO for LES of 3D forced HIT and achieved faster and more accurate prediction compared to the classical LES with the dynamic Smagorinsky model and dynamic mixed model.\cite{li2022fourier} Peng et al. proposed a linear attention coupled Fourier neural operator (LAFNO) to further improve the model accuracy in simulating 3D forced HIT, and free-shear turbulence.\cite{peng2023linear} However, model errors will accumulate over time, leading a challenge for maintaining high accuracy in long-term predictions. In addition, the memory size imposes a limitation on number of layers in the original form of FNO.

In this work, we investigate the effectiveness of implicit layer that utilize shared Fourier layers, which can enable the neural network to be expanded to greater depths, thereby enhancing its capability to approximate complex functions. Simultaneously, we incorporate the U-net network to complement small-scale information, which further enhances the stability of the model. The results demonstrate that the proposed IU-FNO model outperforms the original FNO model in terms of accuracy and stability in predicting 3D turbulent flows, including forced homogeneous isotropic turbulence, free-shear turbulent mixing layer, and decaying homogeneous isotropic turbulence. Moreover, IU-FNO demonstrates long-term stable predictions, which has not been achieved by previous versions of FNO. In comparison with the original FNO, IU-FNO reduces the network parameters by approximately 75\%, and the number of parameters is independent of the number of network layers. Meanwhile, the IU-FNO model also demonstrates improved generalizability to higher Reynolds numbers, and can predict unseen flow regime in decaying turbulence. Therefore, the IU-FNO approach serves as a valuable guide for modeling large-scale dynamics of more complex turbulence. Since we are using a purely data-driven approach without explicitly embedding any physical knowledge, the predicted results might not strictly satisfy the N-S equations. However, the IU-FNO model is capable of approximating the N-S equations from data.

One limitation of the proposed model is that it has only been tested on simple flows, whereas the flows in engineering applications are often much more complex. While the IU-FNO model is effective in predicting flow types under uniform grid and periodic boundary conditions, it requires further improvement to be applicable to non-uniform grid and non-periodic boundary conditions. Another disadvantage of the proposed model is its high dependence on data. As a purely data-driven model, it requires a substantial amount of data for training. Recently, more sophisticated improvements of the FNO framework have been proposed to simulate complex flows, including the adaptive Fourier neural operators (AFNO)\cite{guibas2021adaptive,kurth2022fourcastnet} and physics-informed neural operator (PINO).\cite{li2021physics,rosofsky2023magnetohydrodynamics} Li et al. introduced geo-FNO, a method that can handle PDEs on irregular geometries by mapping the input physical domain into a uniform latent space using a deformation function. The FNO model with the FFT is then applied in the latent space to solve the PDEs.\cite{li2022fouriergeo} The ability to handle arbitrary geometries is essential for solving engineering flows, which often involve complex geometries with irregular boundaries. Most of advanced FNO variants have been only tested in 2D problems, whereas most flows in engineering applications are 3D. In future work, the geo-FNO can be extended and integrated with the proposed IU-FNO models for fast simulations of 3D complex turbulence.

\section{\label{sec:7}Concludsion}
In this work, we proposed an implicit U-Net enhanced Fourier neural operator (IU-FNO) model to predict long-term large-scale dynamics of three-dimensional turbulence. The IU-FNO is verified in the large-eddy simulations of three types of 3D turbulence, including forced homogeneous isotropic turbulence, free-shear turbulent mixing layer, and decaying homogeneous isotropic turbulence.

Numerical simulations demonstrate that: 1) The IU-FNO model performs a superior capability to reconstruct a variety of statistics of velocity and vorticity fields, and the instantaneous spatial structures of vorticity, compared to other FNO-based models and classical DSM model. 2)The IU-FNO model has the capability of accurate predictions for long-time dynamics of 3D turbulence, which can not be achieved by previous forms of FNO. 3) IU-FNO model employs implicit loop Fourier layers to reduce the number of network parameters by approximately 75\% compared to the original FNO. 4)The IU-FNO model is much more efficient than traditional LES with DSM model, and shows an enhanced capacity for generalization to high Reynolds numbers, and can make predictions on unseen flow regime of decaying turbulence. Therefore, the proposed IU-FNO approach has the great potential in developing advanced neural network models to solve 3D nonlinear problems in engineering applications.

\begin{acknowledgments}
This work was supported by the National Natural Science Foundation of China (NSFC Grant Nos. 91952104, 92052301, 12172161 and 91752201), by the NSFC Basic Science Center Program (grant no. 11988102), by the Shenzhen Science and Technology Program (Grants No.KQTD20180411143441009), by Key Special Project for Introduced Talents Team of Southern Marine Science and Engineering Guangdong Laboratory (Guangzhou) (Grant No. GML2019ZD0103), and by Department of Science and Technology of Guangdong Province (No.2020B1212030001). This work was also supported by Center for Computational Science and Engineering of Southern University of Science and Technology.
\end{acknowledgments}

\section*{AUTHOR DECLARATIONS}
\subsection*{Conflict of Interest}
The authors have no conflicts to disclose.

\section*{Data Availability}
The data that support the findings of this study are available from the corresponding author upon reasonable request.

\appendix
\section{\label{app:LES}Conventional subgrid-scale model for LES}
In this appendix, we mainly introduce three classical LES models, including dynamic Smagorinsky model (DSM), velocity gradient model (VGM) and dynamic mixed model (DMM). One of the most widely used functional models is the Smagorinsky model, given by\cite{smagorinsky1963,lilly1967,germano1992}
\begin{equation}
	\tau_{i j}-\frac{\delta_{i j}}{3} \tau_{k k}=-2 C_s^2 \Delta^2|\bar{S}| \bar{S}_{i j},	
	\label{eq7}
\end{equation}
where $\bar{S}_{ij}=\frac{1}{2}\left(\partial \bar{u}_i / \partial x_j+\partial \bar{u}_j / \partial x_i\right)$ represents the strain rate of the filtered velocity, and $|\bar{S}|=\left(2 \bar{S}_{i j} \bar{S}_{i j}\right)^{1 / 2}$ stands for the characteristic filtered strain rate. $\delta_{i j}$ is the Kronecker delta operator, and $\Delta$ denotes the filter width. 

The coefficient $C_s^2$ can be obtained either through theoretical analysis or empirical calibration.\cite{lilly1967} The widely adopted strategy involves implementing the least-squares dynamic methodology by utilizing the Germano identity, resulting in the dynamic Smagorinsky model (DSM) with the coefficient given by\cite{germano1991,lilly1992}
\begin{equation}
	C_s^2=\frac{\left\langle \mathcal{L}_{i j} \mathcal{M}_{i j}\right\rangle}{\left\langle\mathcal{M}_{k l} \mathcal{M}_{k l}\right\rangle} .	
	\label{eq8}
\end{equation}
Here, the Leonard stress $\mathcal{L}_{i j}={\widetilde{\bar{u}_i \bar{u}}_j}-\tilde{\bar{u}}_i \tilde{\bar{u}}_j$, and $\mathcal{M}_{i j}=\tilde{\alpha}_{i j}-\beta_{i j}$ with $\alpha_{i j}=2 \Delta^2|\bar{S}| \bar{S}_{i j}$ and $\beta_{i j}=2 \tilde{\Delta}^2|\tilde{\bar{S}}| \tilde{\bar{S}}_{i j}$. Specially, an overbar denotes the filtering at scale $\Delta$, a tilde represents the test filtering operation at the double-filtering scale $\tilde{\Delta}=2 \Delta$.

A representative structural model is the velocity gradient model (VGM) based on the truncated Taylor series expansions, given by\cite{clark1979evaluation}
\begin{equation}
		\tau_{i j}=\frac{\bar{\Delta}^2}{12} \frac{\partial \bar{u}_i}{\partial x_k} \frac{\partial \bar{u}_j}{\partial x_k} .	
		\label{eqVGM}
\end{equation}

The dynamic mixed model (DMM) combines the scale-similarity model with the dissipative Smagorinsky term, and is given by\cite{liu_meneveau_katz_1994,shi2008constrained}
\begin{equation}
		\tau_{i j}=C_1 \bar{\Delta}^2|\bar{S}| \bar{S}_{i j}+C_2\left(\widetilde{\bar{u}_i \bar{u}_j}-\tilde{\bar{u}}_i \tilde{\bar{u}}_j\right) .
		\label{eqDMM}
\end{equation}
Here, an overbar denotes the filtering at scale $\Delta$, and a tilde represents the test filtering operation at the double-filtering scale $\tilde{\Delta}=2 \Delta$. The spectral filter is employed to double-filtering in HIT, and a Gaussian filter is utilized in free-shear turbluent mixing layer. Similar to the DSM model, the model coefficients $C_1$ and $C_2$ of the DMM model are dynamically determined using the Germano identity through the least-squares algorithm. $C_1$ and $C_2$ are expressed as\cite{yuan2020deconvolutional,xie2020artificial}
\begin{equation}
		C_1=\frac{\left\langle N_{i j}^2\right\rangle\left\langle L_{i j} M_{i j}\right\rangle-\left\langle M_{i j} N_{i j}\right\rangle\left\langle L_{i j} N_{i j}\right\rangle}{\left\langle N_{i j}^2\right\rangle\left\langle M_{i j}^2\right\rangle-\left\langle M_{i j} N_{i j}\right\rangle^2} ,
		\label{eqC1}
\end{equation}
\begin{equation}
		C_2=\frac{\left\langle M_{i j}^2\right\rangle\left\langle L_{i j} N_{i j}\right\rangle-\left\langle M_{i j} N_{i j}\right\rangle\left\langle L_{i j} M_{i j}\right\rangle}{\left\langle N_{i j}^2\right\rangle\left\langle M_{i j}^2\right\rangle-\left\langle M_{i j} N_{i j}\right\rangle^2} ,
		\label{eqC2}
\end{equation}
	where $M_{ij}=H_{1, i j}-\tilde{h}_{1, i j}$, and $N_{i j}=H_{2, i j}-\tilde{h}_{2, i j}$. Here, $h_{1, i j}=-2 \bar{\Delta}^2|\bar{S}| \bar{S}_{ij}$, $h_{2,ij}=\widetilde{\bar{u}_i \bar{u}_j}-\tilde{\bar{u}}_i \tilde{\bar{u}}_j$, $H_{1, i j}=-2 \tilde{\Delta}^2|\tilde{\bar{S}}| \tilde{\bar{S}}_{i j}$, and $H_{2, i j}=\widehat{\tilde{\bar{u}}_{i} \tilde{\bar{u}}_{j}} -\hat{\tilde{\bar{u}}}_i \hat{\tilde{\bar{u}}}_j$ The hat denotes the filter at scale $\hat{\Delta}=4\Delta$.

\section{\label{app:NN}The Details of related Fourier neural operator methods}
In this appendix, we introduce the details of related Fourier neural operators, including U-FNO and IFNO.

\begin{figure*}
	\includegraphics[width=1\linewidth]{ 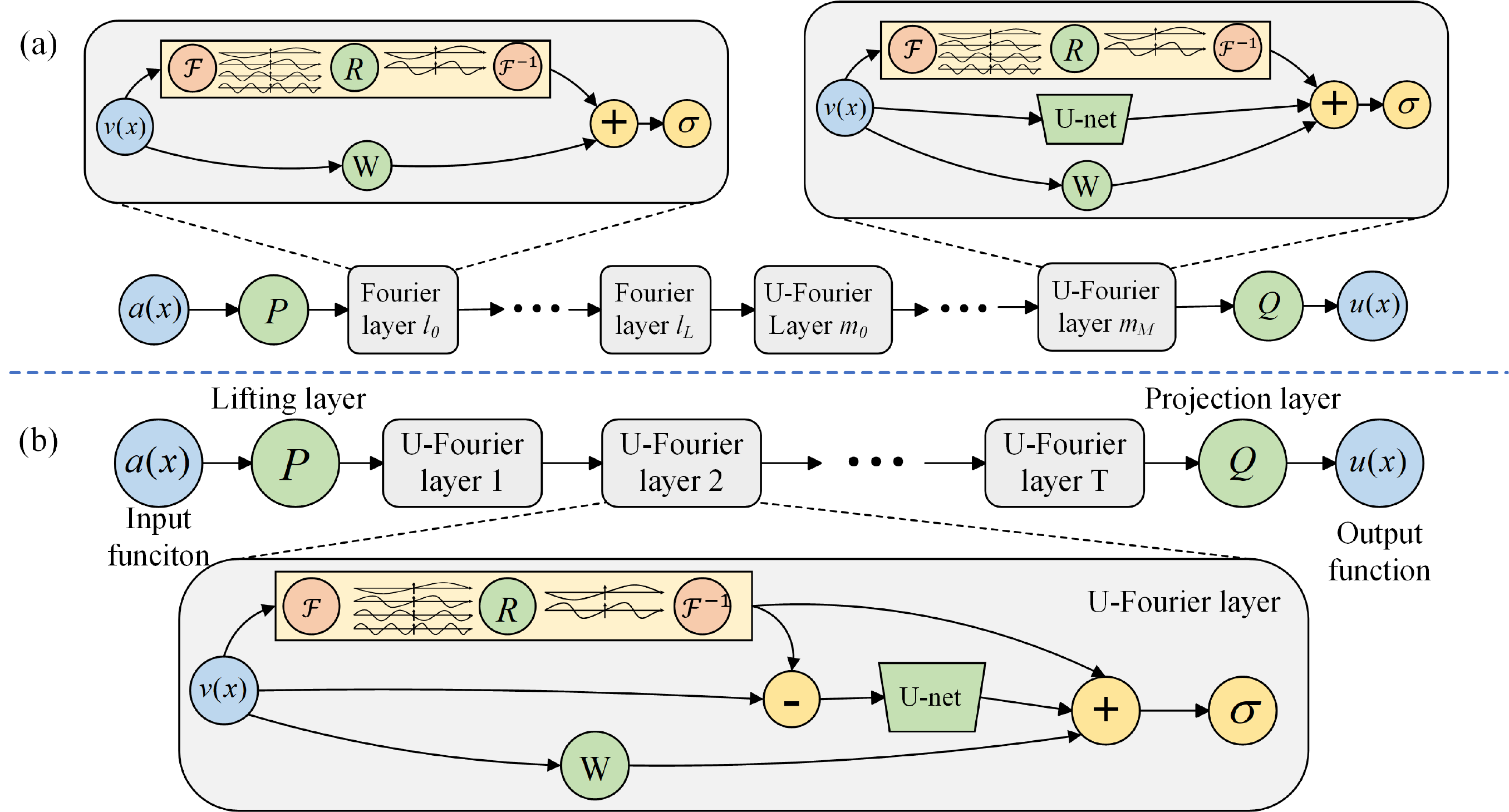}
	\caption{The architectures of U-Net enhanced Fourier neural operators (U-FNO).(a) the architecture of U-FNO proposed by Wen et al.\cite{wen2022u} (b) the architecture of modified U-FNO proposed by us.}
	\label{NNUFNO}
\end{figure*}

Fig.~\ref{NNUFNO}(a) illustrates the architectures of the U-FNO model. The U-FNO employ iterative architectures: $v_{l_{0}} \mapsto v_{l_{1}} \mapsto \ldots \mapsto v_{l_{T}} \mapsto v_{m_{0}} \ldots \mapsto v_{m_{M}}$ where $v_{l_{j}}$ for $j=0,1, \ldots, T-1$ and $v_{m_{k}}$ for $k=0,1, \ldots, M-1$ are sequences of functions taking values in $\mathbb{R}^{d_v}$.\cite{wen2022u} Therefore, when doing local transformation projection, the operation $u(x)=$ $Q\left(v_{m_{M}}(x)\right)$ is performed on $v_{m_{M}}(x)$. Specifically, $v_{l_{j}}$ denotes $j$-th Fourier layer which is the same as original FNO architecture. $v_{m_{M}}$ represents $M$-th U-Fourier layer which is given as
\begin{equation}
	v_{m_{k+1}}(x):=\sigma\left(\mathcal{F}^{-1}\left(R_\phi \cdot\left(\mathcal{F} v_{m_k}(x)\right)\right)(x)+\left(\mathcal{U^*} v_{m_k}\right)(x)+W\left(v_{m_k}(x)\right)\right),  \quad \forall x \in D .
	\label{eq13}
\end{equation}
Here, $\mathcal{F}$, $\mathcal{F}^{-1}$, $R_\phi$ and $W$ have the same meaning as defined in Section~\ref{sec:3.1}. $\mathcal{U}^*$ denotes a U-Net CNN operator. The architecture of U-FNO differs from the original Fourier layer in FNO by incorporating a U-Net path into each U-Fourier layer. The purpose of the U-Net is to perform local convolutions that enhance the representation capability of the U-FNO, particularly for small-scale flow structures.

\begin{figure*}
	\includegraphics[width=1\linewidth]{ 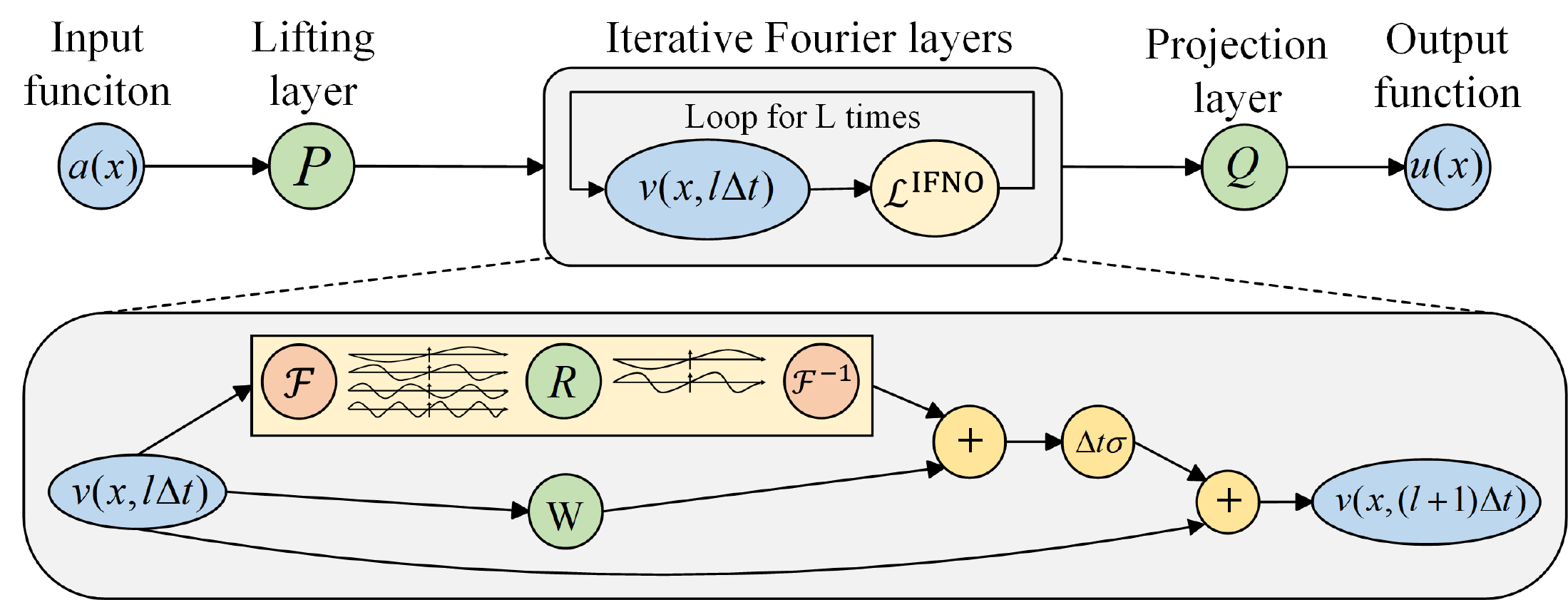}
	\caption{The architecture of implicit Fourier neural operator (IFNO).}
	\label{NNIFNO}
\end{figure*}
The architecture of IFNO is shown in Fig.~\ref{NNIFNO}, which can greatly reduce the number of trainable parameters and memory cost, and overcome the vanishing gradient problem of training networks with deep layers. The iterative network update of IFNO is given as\cite{you2022learning}
\begin{equation}
	\begin{aligned}
		{v}(x,(l+1) \Delta t) & =\mathcal{L}^{I F N O}[{v}({x}, l \Delta t)] \\
		& :={v}({x}, l \Delta t)+\Delta t \sigma\left(W {v}({x}, l \Delta t)+ \mathcal{F}^{-1}\left(R_\phi \cdot\left(\mathcal{F} {v}({x}, l \Delta t)\right)\right)(x) \right), \forall x \in D. 
	\end{aligned}
	\label{eq16}
\end{equation}

The IFNO model employs a parameter-sharing strategy and continuously optimizes the network parameters through iterative loops to enhance its accuracy. This approach is effective in improving the performance of the network and enables it to handle complex data and nonlinear problems more efficiently.

\bibliography{aipsamp}

\end{document}